\documentclass[11pt,a4paper]{article}
\pdfoutput=1
\usepackage{jheppub}
\usepackage{amsmath,amssymb}
\usepackage{graphicx}
\usepackage{dcolumn}
\usepackage{bm}
\usepackage{cancel}
\usepackage{scrextend}
\usepackage{slashed}

\newcommand{\doublet}{SU(3)/[SU(2)\times U(1)]}
\newcommand{\triplet}{[SU(2)^2\times U(1)]/[SU(2)\times U(1)]}
\newcommand{\mchm}{SO(5)/SO(4)}

\newcommand\w[1]{\makebox[1em]{$#1$}}

\begin{document}

\preprint{DESY 15-047}

\title{Composite Dark Sectors}

\date{\today}

\author[a]{Adri\'an Carmona}
\author[b]{and Mikael Chala}
\affiliation[a]{Institute for Theoretical Physics, ETH Zurich, 8093 Zurich, Switzerland}
\affiliation[b]{DESY, Notkestra$\ss$e 85, 22607 Hamburg, Germany}
\emailAdd{carmona@itp.phys.ethz.ch}
\emailAdd{mikael.chala@desy.de}

\abstract{We introduce a new paradigm in Composite Dark Sectors, where the full Standard Model (including the Higgs boson) is extended with a strongly-interacting composite sector with global symmetry group $\mathcal{G}$ spontaneously broken to $\mathcal{H}\subset \mathcal{G}$. We show that, under well-motivated conditions, the lightest neutral pseudo Nambu-Goldstone bosons are natural dark matter candidates for they are protected by a parity symmetry not even broken in the electroweak phase. These models are characterized by only two free parameters, namely the typical coupling $g_D$ and the scale $f_D$ of the composite sector, and are therefore very predictive. We  consider in detail two minimal scenarios, $\doublet$ and $\triplet$, which provide a dynamical realization of the Inert Doublet and Triplet models, respectively. We show that the radiatively-induced potential can be computed in a five-dimensional description with modified boundary conditions with respect to Composite Higgs models.  Finally, the dark matter candidates are shown to be compatible, in a large region of the parameter space, with current bounds from dark matter  searches as well as electroweak and collider constraints on new resonances.}

\maketitle

\section{Introduction}

The Standard Model (SM) of elementary particles has been tested to an impressive level of accuracy in a large amount of experiments. In particular, the current Large Hadron Collider (LHC) data do not reveal any significant departure from the SM predictions, leaving the experimental evidence of new physics mainly in the non-vanishing neutrino masses and the Dark Matter (DM) observations. The concrete nature of the new physics is however still a matter of study. Nevertheless, concerning DM, a weak interacting massive particle (WIMP) has long been a prime candidate. Furthermore, in light of the celebrated experimentally established Higgs sector, scalar WIMPs have earned some attention. The addition of a singlet elementary scalar, for instance, has been detailed considered in the recent years (see for example Refs.~\cite{Djouadi:2011aa,Mambrini:2011ri,Kadastik:2011aa,Batell:2011pz,He:2011gc}). More elaborated models comprise extensions with composite scalars. Among them, composite Higgs models (CHMs) deserve special attention, inasmuch as they provide an appealing alternative to supersymmetry. In CHMs with DM candidates, both the Higgs and the WIMP particles are assumed to be pseudo-Nambu-Goldstone bosons (pNGBs) of a new strongly-interacting sector, with a global symmetry
breaking pattern $\mathcal{G}/\mathcal{H}$. Hence, they can be naturally much lighter than the new physics compositeness scale. Explicit models have been worked out for instance in Refs.~\cite{Frigerio:2012uc, Gripaios:2009pe, Chala:2012af, Barnard:2014tla}.

In this article we want to address the phenomenological implications of a related but different scenario. We consider SM extensions in which \textit{only} the DM particles are in the pNGB composite sector, being the SM matter content (including the Higgs boson, $H$) fully elementary. (Similar approaches have been previously considered in the literature. See e.g.~\cite{Kilic:2009mi,Kilic:2010et,Antipin:2014qva,Antipin:2015xia}). The global symmetry of the composite sector is therefore only broken by the SM gauge interactions. As we shall detail below, this class of models presents several interesting features: \textit{(i)} if the coset is symmetric (see footnote~\ref{foot:symmetric}) the DM candidate is protected by a parity symmetry, not even broken by the loop-induced Coleman-Weinberg potential; \textit{(ii)} this class of models turns out to be extremely predictive, for there are only two free parameters involved, corresponding to the scale of compositeness $f_D$ and the typical coupling $g_D$ of the DM sector; and \textit{(iii)} in the simplest scenarios, the values of $g_D$ and $f_D$ reproducing the relic abundance make the model evade current constraints while being under the reach of future experiments. In the following, we will discuss the  general structure of these models, while detailed calculations will be provided for two minimal cosets, namely $\triplet$ and $\doublet$.

The article is structured as follows. In section~\ref{sec:chms} we describe the general properties of this scenario, and classify the smallest realizations that can be considered distinguishing those phenomenologically viable. Based on purely symmetry and scale arguments, we estimate the size of the couplings, masses and hence the expected phenomenology on the minimal setups. We write the complete phenomenologically relevant Lagrangian in section~\ref{sec:potential}. In that section the Coleman-Weinberg potential is also computed in a holographic framework. Special attention is paid to the way in which this formalism, originally describing CHMs, is redefined. Indeed, this new setup requires the boundary conditions in the ultraviolet (UV) brane to be properly modified. In section~\ref{sec:pheno}, we consider current bounds. We show that these are dominated by relic-abundance measurements and LHC searches of long-lived charged particles and heavy resonances decaying into pairs of SM fermions. We conclude in section~\ref{sec:conclusions}.

\section{A different look at Composite Models}
\label{sec:chms}

\begin{figure}[t]
	\begin{center}
\includegraphics[width=0.75\textwidth]{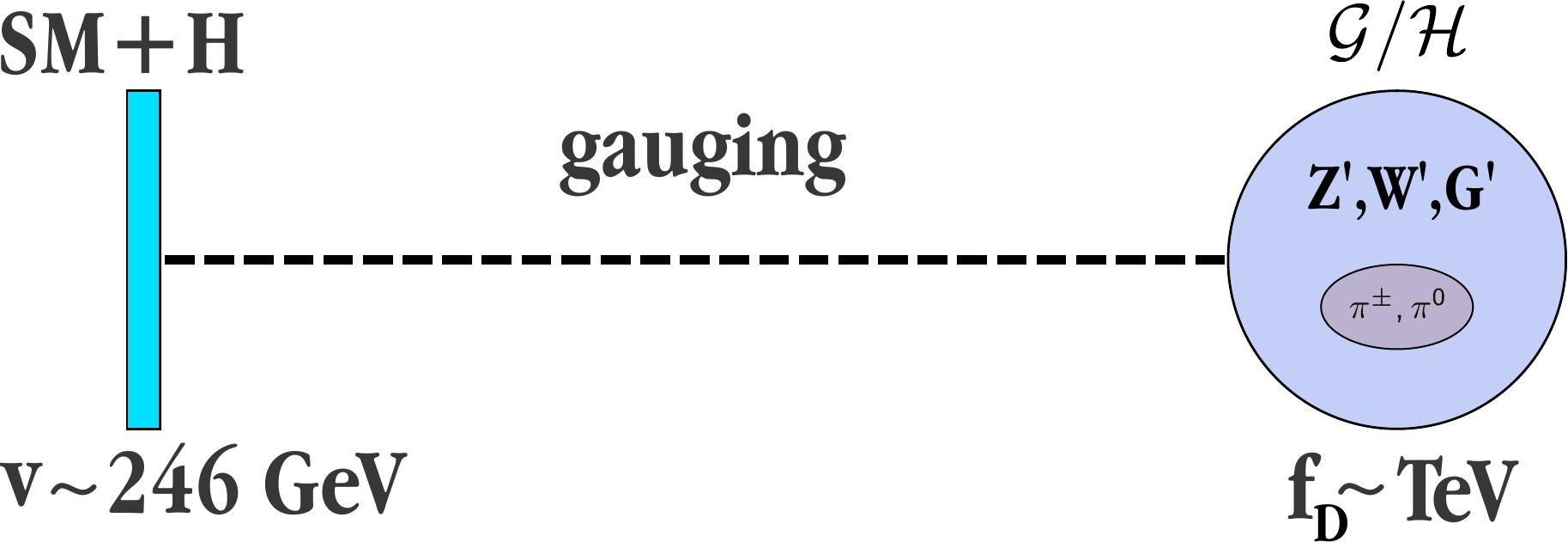}
\end{center}
\caption{Pictorial representation of a Composite Dark Sector. The SM gauge bosons are the only responsible for the interaction between the elementary sector (including the Higgs boson) and the composite sector.}
\label{fig:framework}
\end{figure}

CHMs~\cite{Terazawa:1976xx, Terazawa:1979pj, Kaplan:1983fs,Kaplan:1983sm,Dimopoulos:1981xc} were originally proposed as a compelling solution to the hierarchy problem. The Higgs boson arises as a bound state of a new strongly-interacting sector, with a global symmetry group $\mathcal{G}$ spontaneously broken to $\mathcal{H} \subset \mathcal{G}$. Thus, its mass is protected by its finite size. The Higgs boson is assumed to be a (pNGB) of the global symmetry breaking pattern. Hence, the Higgs boson mass can be naturally at the electroweak (EW) scale if the new physics scale $f$ is around the TeV. The symmetry $\mathcal{G}$ is explicitly broken in two ways: by the gauging of only the SM gauge subgroup $\mathcal{G}_{\rm SM}$ of $\mathcal{H}$, and by the linear mixing of the composite resonances with the elementary SM fields. As a consequence, the physical fields are admixtures of composite and elementary states (partial compositeness~\cite{Kaplan:1991dc}). Then, the larger the mixing, the larger the interaction of an elementary field with the strong sector ---in particular with the Higgs boson, which is fully composite--- and hence its mass. Thus, contrary to what happens in the SM, the Higgs potential is dynamically generated. However, as first stated by Witten~\cite{Witten:1983ut}, the radiative contribution from gauge fields generate a potential whose vacuum expectation value (VEV) is aligned in the direction that preserves the gauge symmetry. Hence, the linear mixing between the elementary fermions and the composite resonances is not only a requirement to correctly reproduce the fermion masses, but also to achieve a realistic EWSB pattern. However, both EW precision data (EWPD) \cite{Grojean:2013qca} and the recent LHC searches  \cite{ATLAS:2015fka} are getting more and more in conflict with natural arguments \cite{Matsedonskyi:2012ym, Marzocca:2012zn, Redi:2012ha, Pomarol:2012qf, Panico:2012uw, Carmona:2014iwa}. Indeed, current analyses are already excluding the verge of the natural partner parameter space.~\footnote{Recently, different models have been proposed to sensibly relax this tension. See for instance Ref.~\cite{Carmona:2014iwa}.} Our aim here is to point out that, in light of the previous discussion, this mechanism can be used not to provide a realistic CHM, but a viable DM explanation. To be concrete, we extend the SM with a strongly-interacting sector with \textit{symmetric}~\footnote{\label{foot:symmetric}By \textit{symmetric coset}, we mean that in which the broken generators, $X$, commute as $[X^i, X^j] = i f_{ijk} T^k$, where $T$ stands for the unbroken generators.} coset $\mathcal{G}/\mathcal{H}$. The SM, including the Higgs boson, is considered to be completely elementary, while we assume that only the SM gauge interactions are the responsible for the breaking of the global symmetry in the composite sector. We will prove that the neutral pNGBs in this scenario (see figure~\ref{fig:framework}) are natural DM candidates.
Indeed, if $\mathcal{G}/\mathcal{H}$ is symmetric, only terms with an even number of pNGBs are present in the non-linear description of the new strong sector. That means that the composite sector automatically respects a $Z_2$ symmetry for which the pNGBs $\pi$ transform as $\pi\rightarrow -\pi$. Clearly, the gauging of $\mathcal{G}_{\rm SM}\subset \mathcal{H}\subset \mathcal{G}$ breaks explicitly the global symmetry, but keeps $Z_2$ conserved. Hence, given that fermions are assumed to be fully elementary, the parity symmetry is not explicitly broken. Not even spontaneously broken, inasmuch as the VEV in the loop-generated Coleman-Weinberg potential is expected to live in the EW-preserving direction, according to the Witten result mentioned above. Furthermore, the pNGB masses are not restricted by experimental searches to be as low as in the Higgs case, what allows the composite resonances to be  heavy enough not to conflict with current constraints. As a matter of fact, non-singlet scalar DM particles have been found to predict the correct relic abundance specially for large masses (see for example Refs.~\cite{Cirelli:2007xd, FileviezPerez:2008bj, Goudelis:2013uca}). These results will be explicitly stated in the next section. All these features make this framework a promising scenario for DM. Before going further, however, let us classify the possible models fitting these requirements, from the simplest to more involved realizations.~\footnote{Hereafter, both $\mathcal{H}$ and $\mathcal{G}$ are always assumed to be multiplied by the unbroken color group $SU(3)$ too, which is only omitted in the equations for a simpler reading. Although its inclusion could be also disregarded, we decided to take it into account since it just leads to more conservative bounds from direct detection. This argument also applies to other spectator groups to first approximation.}
\begin{itemize}
	\item $\mathcal{H} = SU(2)\times U(1)$, $\mathcal{G} = SU(2)\times U(1)^2$. In this case, the broken $U(1)$ commutes with the whole SM gauge group. Thus, gauge-boson loops do not generate a potential for the corresponding neutral pNGB, for it is not charged under the SM gauge symmetry. As a consequence, it is an exact NGB and hence massless. This simplest model should be therefore disregarded from the phenomenological point of view.

	\item $\mathcal{H} = SU(2)\times U(1)$, $\mathcal{G} = SU(2)\times U(1)^n$. For any $n$, the argument above still applies.

	\item $\mathcal{H} = SU(2)\times U(1)$, $\mathcal{G} = SU(2)^2$. In this case, only two \textit{charged} states emerge. In light of the strong bounds on stable charged particles~\cite{Goudelis:2013uca,Dimopoulos:1989hk,Chivukula:1989cc,Gould:1989gw}, this kind of models are not phenomenologically viable. More in general, pNGB spectra with only neutral fields are not expected. Consequently, the minimal realistic models do have at least three degrees of freedom, corresponding to one neutral and two charged states.

	\item $\mathcal{H} = SU(2)\times U(1)$, $\mathcal{G} = SU(2)^2\times U(1)$. This is the simplest realistic realization. A real scalar triplet with hypercharge $Y = 0$ appears in the spectrum. We will analyze this model in detail in next sections.

	\item $\mathcal{H} = SU(2)\times U(1)$, $\mathcal{G} = SU(2)^2\times U(1)^n$. As in the first case, massless scalars are expected. Thus, we no longer consider this coset. Besides, further extensions with more $SU(2)$ insertions give rise to coset spaces larger than the ones considered below.

	\item $\mathcal{H} = SU(2)\times U(1)$, $\mathcal{G} = SU(3)$. This symmetry-breaking pattern has been previously considered in the CHM literature (see for instance Ref.~\cite{Contino:2003ve}). However, it has been disregarded in the recent years for it gives large corrections to the $\rho\approx 1$ parameter (it does break the custodial symmetry in the Higgs sector). In our setup, instead, the Higgs boson lives in the elementary sector and thus we will also take this coset into account.

\end{itemize}
Larger group combinations either provide cosets with more degrees of freedom or scenarios phenomenologically similar to the ones we have enumerated previously. A prime example is $\mchm$. This coset structure stands for the minimal CHM and has been largely studied in the literature~\cite{Agashe:2004rs}. This model has the advantage of containing no anomalous representations, that potentially distort the parity symmetry~\cite{Gripaios:2009pe}. At any rate, in the following we only consider in detail the cosets $\doublet$ and $\triplet$, assuming that their UV completions make them anomaly-free. Finally, a last comment is in order. Dark multiplets with fractional hypercharge have been typically ignored so far. The reason is that these multiplets are always complex, and so two neutral states are predicted. The problem arises because these states are generically degenerated in mass, and they are hence strongly constrained by direct detection experiments~\cite{Hambye:2009pw}. The exception is the well-known case where a second EW doublet $\Phi$ is added, the so-called Inert Doublet Model (IDM) \cite{Deshpande:1977rw, Ma:2006km, Barbieri:2006dq}. In this case, the operator $\lambda\,[(H^\dagger\Phi)^2+\text{h.c.}]$ can be written at the tree level. After EWSB, this operator introduces a small splitting in the mass squared of order $\approx\lambda v^2$ between the two neutral components. This operator, however, does not arise at the quantum level in the Composite Dark Sector scenario provided by $\doublet$. The reason is that, in the massless limit, the pNGB sector respects a $U(1)$ symmetry (that includes the $Z_2$ parity) under which $\Phi\rightarrow \exp{(i\theta)}\Phi$, being this symmetry still exact after gauging the SM gauge group. At any rate, it can be always assumed this symmetry to be broken at a higher scale. Not only in the $Y = 1/2$ case, but also for larger quantum numbers. As a matter of fact, for any $SU(2)_L$ multiplet $\phi$ with half-integer hypercharge  and isospin $I$ satisfying the relation $Y=I$, this splitting can be achieved by a higher-dimensional operator with $m=2 I$ Higgs insertions and an arbitrary number $n$ of Higgs singlet operators $\lambda[(H^\dagger H)^n(\tilde{H}\tilde{H}\overset{m}{\cdots}\tilde{H}\phi)^2+\text{h.c.}]$, where $\tilde{H}=i\sigma^2 H^{\ast}$. Note that a new physics scale is anyway required in order to stabilize the Higgs mass. Hereafter we assume that this is the case in the coset $\doublet$. Provided the splitting is small (as it can be expected if the additional $U(1)$ symmetry is broken at a large scale), the dependence of other constraints as well as future searches on it is negligible. Thus, the model is effectively still described by just two parameters.

A first good approximation to the size of the masses and couplings of the degrees of freedom that arise in both $\triplet$ and $\doublet$ can be obtained considering the symmetries and scales involved in this framework. Indeed, given that the new sector is assumed to be strongly coupled, $g_D$ is expected to be $1\lesssim g_D\lesssim 4\pi$. Thus, the new resonances are expected to have a mass $m_\rho\approx g_D f_D\approx $ few TeV in order not to conflict with current constraints. Thus, the pNGBs $\pi$ (which are massive only because of the explicit breaking of the global symmetry) are naturally expected to live around (or slightly below) $\approx 1$ TeV. In addition, the non-derivative scalar interactions, being generated by loops of SM gauge bosons, are predicted to be sub-dominant with respect to the gauge interactions. Besides, as both the electromagnetic charged and neutral pNGBs come in complete representations of the SM gauge group, the mass difference among themselves can only arise in the EW broken phase, being hence proportional to some power of the EW VEV $v\approx 246$ GeV. (As a matter of fact, there is only one renormalizable operator that can break the degeneracy after EWSB, given by $(H^\dagger\tau^a H)M_{ab}(\pi^\dagger \Gamma^b\pi)$, where $\tau$ and $\Gamma$ are the Pauli matrices and the $SU(2)$ generators in the representation $\pi$ belongs to in the spherical basis,~\footnote{The Pauli matrices in the spherical basis are written as $\tau^{\pm 1} = \pm(\sigma^1\mp i\sigma^2)/2$ and $\tau^0 = \sigma^3/2$. The same applies for larger representations.} and $M_{ab}$ is the three times three anti-diagonal matrix $\text{adiag} (1, -1, 1)$.) All together implies a small splitting between the neutral and the charged states, that makes the latter typically long-lived. As we shall discuss in section~\ref{sec:pheno}, searches of these particles in the current LHC data~\cite{Chatrchyan:2013oca,ATLAS:2014fka} provide one of the strongest constraints in some scenarios. 

\section{Gauge interactions, scalar potential and masses} 
\label{sec:potential}

Let us consider the two cosets $\triplet$ and $\doublet$. The group matrices are explicitly written in the Appendices~\ref{app:triplet} and~\ref{app:doublet} respectively. The dynamics of the composite DM boson at low energy is described by a non-linear sigma model, with non-linear interactions parameterized by the scale $f_D$. Following the CCWZ formalism~\cite{Coleman:1969sm,Callan:1969sn}, the derivative and gauge interactions among the pNGBS are given by the trace of $d_\mu d^\mu$, where $d_\mu$ stands for the perpendicular projection of the Lie-algebra-valued Cartan one-form $\omega_\mu = -iU^\dagger D_\mu U = d^a_\mu X^a + E_\mu^a T^a$, with $U=\exp{(-i\sum \pi^a X^a/f_D)}$, $X^a$ the broken generators, $T^a$ the unbroken ones and $D_\mu$ the covariant derivative:
\begin{equation}
D_\mu\varphi = \bigg(\partial_\mu+ig_s\frac{\lambda_a}{2}g_\mu^a+igT^IW^I_\mu+ig'YB_\mu\bigg)\varphi,
\end{equation}
which can be also written as
\begin{equation}\label{eq:covariant_derivative}
D_\mu\varphi = \bigg[\partial_\mu+ig_s\frac{\lambda_a}{2}g_\mu^a+\frac{ig}{\sqrt{2}}\left(T^+W_\mu^++T^- W_\mu^-\right)+\frac{ig}{c_W}\left(T_3-s_W^2Q\right)Z_\mu + ieQA_\mu\bigg]\varphi,
\end{equation}
if we define $T^{\pm} = T^1\pm iT^2$ and $W^\pm = (W_\mu^1\mp iW_\mu^2)/\sqrt{2}$. We have denoted by $s_W$ ($c_W$) the sine (cosine) of the SM Weinberg angle $\theta_W$ and $e$ is the electromagnetic unit charge. For $\triplet$ the relevant Lagrangian~\footnote{For the computation of the potential (see next section) all the terms with larger powers of $1/f_D$ are included. So, once the whole potential is known, the study of the DM and the LHC phenomenology can be completed with the addition of the Lagrangian above.} reads:
\begin{align}\label{eq:triplet}
\mathcal{L} &= g^2 (\pi^0)^2 W^+_\mu W^{\mu\,-} + \bigg[i g W^{\mu\,+} (\pi^0 \overleftrightarrow{\partial_\mu}\pi^-) -\frac{1}{2}g^2 W^+_\mu W^{+\,\mu} \pi^-\pi^- +\text{h.c.}\bigg]\\\nonumber
&+g^2W^+_\mu W^{\mu\,-}\pi^+\pi^-+\frac{g^2}{c_W^2} (s_W^2-1)^2 Z_\mu Z^\mu \pi^+\pi^- + \frac{ig(1-s_W^2)}{c_W} Z^\mu(\pi^+\overleftrightarrow{\partial_\mu}\pi^-)\\\nonumber
&+ e^2 A_\mu A^\mu \pi^+ \pi^- + ie A^\mu (\pi^+\overleftrightarrow{\partial_\mu}\pi^-)+\frac{2eg}{c_W}(s_W^2-1)A_\mu Z^\mu\pi^+\pi^-\\
&+ \bigg[eg A_\mu\pi^0W^{\mu\,+}\pi^- + \frac{g^2}{c_W}(s_W^2-1) W^+_\mu Z^\mu \pi^0  \pi^- + \text{h.c.}\bigg] + \frac{1}{2f_D^2}[\partial_\mu(\pi^0)^2]\partial^\mu(\pi^+\pi^-).\nonumber
\end{align}
For $\doublet$ it reads
\begin{align}\label{eq:doublet}
	\mathcal{L} &= \frac{g^2}{4} |\Pi^0|^2 W^+_\mu W^{\mu\,-} + \bigg[\frac{i g}{2} W^{\mu\,+} (\Pi^0 \overleftrightarrow{\partial_\mu}\pi^-) -\frac{1}{2}g^2 W^+_\mu W^{\mu\,+} \pi^-\pi^- +\text{h.c.}\bigg]\\\nonumber
&+\frac{g^2}{2}W^+_\mu W^{\mu\,-}\pi^+\pi^-+\frac{g^2}{4c_W^2} (1-2s_W^2)^2 Z_\mu Z^\mu \pi^+\pi^- + \frac{ig(1-2s_W^2)}{2c_W} Z_\mu(\pi^+\overleftrightarrow{\partial_\mu}\pi^-)\\\nonumber
&+ e^2 A_\mu A^\mu \pi^+ \pi^- + ie A_\mu (\pi^+\overleftrightarrow{\partial_\mu}\pi^-)+\frac{eg}{c_W}(1-2s_W^2)A_\mu Z^\mu\pi^+\pi^-\\\nonumber
&+ \bigg(\frac{eg}{2} A_\mu\Pi^0W^+_\mu\pi^- - \frac{e^2}{2c_W} W^+_\mu Z^\mu \Pi^0  \pi^- + \frac{ig}{2c_W}\Pi^{0\dagger}\partial_\mu\Pi^0 Z^\mu+\text{h.c}\bigg)\\\nonumber
&+\frac{g^2}{8c_W^2}Z^\mu Z_\mu |\Pi^0|^2+\frac{1}{2f_D^2}[\partial_\mu |\Pi^0|^2]\partial^\mu(\pi^+\pi^-),
\end{align}
 In the equations above, $\Pi^0 = \pi^0+iA^0$, $\pi^0$ and $A^0$ stand for the neutral pNGBs and $\pi^+$ and $\pi^-$ are the charged ones. In addition, we have defined $f\overleftrightarrow{\partial_{\mu}}g=f\partial_{\mu}g-(\partial_{\mu}f) g$. We have omitted higher-dimensional operators, which are suppressed by larger powers of $1/f_D$, with the exception of the last term in eqs.~(\ref{eq:triplet}) and (\ref{eq:doublet}). The reason is that, although a detailed computation of gamma-ray observables is beyond the scope of our study,~\footnote{As well as other relevant gamma-ray spectral features, see e.g. \cite{Gustafsson:2007pc,Garcia-Cely:2013zga} for the IDM case.} it is worth mentioning that these operators can play an important role on such processes. Indeed, processes like $\pi^0 \pi^0\rightarrow \gamma \gamma (Z)$ receive contributions from loop triangles with virtual $\pi^\pm$. The coupling between the neutral and the charged states is usually given by the quartic term $\lambda_{\pi^0\pi^{\pm}}(\pi^0)^2\pi^+\pi^-$ in the potential. However, the contribution from the derivative interaction, $\approx m^2_{\pi^0}/f^2_D$, can easily exceed the coupling $\lambda_{\pi^0\pi^{\pm}}$. Specially in these composite models, in which the latter is radiatively induced.

 In order to estimate the strongly-coupled effects, we work out a holographic scenario described by a modified five-dimensional (5D) description of CHMs, where the full SM lives on the UV brane and just gravity and gauge interactions propagate through the bulk of the extra dimension. In particular, we consider an AdS$_5$ space with metric \cite{Randall:1999ee}
\begin{eqnarray}
	\mathrm{d}s^2=a^2(z)(\eta_{\mu\nu}\mathrm{d}x^{\mu} \mathrm{d}x^{\nu}-\mathrm{d}z^2)= \left(\frac{R}{z}\right)^2(\eta_{\mu\nu}\mathrm{d}x^{\mu} \mathrm{d}x^{\nu}-\mathrm{d}z^2),
\end{eqnarray}
where $z\in [R,R^{\prime}]$ is the coordinate of the additional spatial dimension and $R$ and $R^{\prime}$ are the positions of the UV and the infra-red (IR) brane, respectively. In order to effectively describe the two breaking patterns introduced before, we will extend in the bulk of the extra dimension the $SU(2)_L\times U(1)_Y$ gauge symmetry of the UV brane to $SU(2)_1\times SU(2)_2\times U(1)_Y$ and $SU(3)\times U(1)_X$, respectively, where the additional $U(1)_X$ gauge group has been added to correctly reproduce the observed Weinberg angle. On the other hand, as the IR boundary conditions parametrize the spontaneous symmetry breaking  of the Goldstone symmetry by the strong dynamics \cite{ArkaniHamed:2000ds,Rattazzi:2000hs},  the 5D bulk gauge symmetry will reduce on the IR brane to $SU(2)_L\times U(1)_Y$ and $SU(2)_L\times U(1)_8\times U(1)_X$, respectively. More specifically, we define
\begin{eqnarray}
	W_{M}^i= \frac{1}{\sqrt{2}}\left[L_{M}^i+ R_{M}^i\right], \quad X_{M}^i= \frac{1}{\sqrt{2}}\left[ L_{M}^i- R_{M}^i\right],\quad M=\mu,5,\quad i=1,2,3,
\end{eqnarray}
in the $\triplet$ case, with $L_{M}^i$ and $R_{M}^i$ being the 5D gauge fields associated to $SU(2)_1$ and $SU(2)_2$, respectively. For the $\doublet$ case, instead, we have
\begin{eqnarray}
	B_{M}&= &s_{\phi}\sqrt{3}W_{M}^8+c_{\phi} X_{M},\qquad Z_{M}^{\prime}= c_{\phi} \sqrt{3}W_{M}^8-s_{\phi} X_{M},\qquad M=\mu,5,\nonumber\\
	c_{\phi}&= &\frac{g_5}{\sqrt{g_5^2+g_X^{2}}},\qquad \ \ \ \ \quad \quad ~ ~ s_{\phi}= \frac{g_X}{\sqrt{g_5^2+g_X^{2}}},
\end{eqnarray}
where $g_5$ and $g_X$ are the dimensionful 5D gauge couplings of $SU(3)$ and $U(1)_X$, while $W_{M}^{I},~ I=1,\ldots,8,$ and $X_{M}$ are the corresponding 5D gauge fields. In each case, the corresponding boundary conditions on the UV before EWSB read~\footnote{We will assume for the moment that the EWSB pattern is the usual one, checking explicitly later that this is indeed the case.}  
\begin{eqnarray}
	W_{\mu}^{i}(+,+),\quad X_{\mu}^{i}(-,-),\quad B_{\mu}(+,+), \quad i=1,2,3, 
\end{eqnarray}
and
\begin{eqnarray}
	W_{\mu}^{i}(+,+),\quad W_{\mu}^{\bar{a}}(-,-),\quad B_{\mu}(+,+),\quad Z_{\mu}^{\prime}(-,+), \quad i=1,2,3,\quad \bar{a}=4,5,6,7,
\end{eqnarray}
respectively. In the previous equations, the first (second) $+/-$ denotes Neumann/Dirichlet boundary conditions at the UV (IR) brane and we have omitted those of the four-dimensional scalar counterparts $(\mu\to 5)$, which have opposite boundary conditions. For both cosets, the gauge fields with UV Neumann boundary conditions, $W_{\mu}^i$ and $B_{\mu}$, will be associated with the SM-like EW gauge bosons, whereas the scalar components of the ones having  Dirichlet boundary conditions at both branes will provide the corresponding Goldstone degrees of freedom
\begin{eqnarray}
	X_5^{i}(x,z)=f(z)\pi^{i}(x)+\ldots, \quad W_{5}^{\hat{a}+3}=f(z)\pi^{\hat{a}}(x)+\ldots, \quad i=1,2,3,\quad \hat{a}=1,2,3,4, \qquad~
\end{eqnarray}
where the dots stand for non-physical Kaluza-Klein (KK) resonances and \cite{Contino:2003ve}
\begin{eqnarray}
	f(z)= a^{-1}(z)\left[\int_R^{R^{\prime}}\mathrm{d}z^{\prime}~a^{-1}(z^{\prime})\right]^{-1/2}.
\end{eqnarray}

Since in these scenarios, and contrary to the usual CHMs, the Higgs doublet is localized on the UV brane, the boundary conditions after EWSB at the UV brane for the gauge bosons with Neumann boundary conditions will be modified. It is thus convenient to define the usual physical combinations $W_{M}^{\pm}= (W^1_M\mp i W^2_M)/\sqrt{2}$ and
\begin{eqnarray}
	A_{M}&= &\hat{s}_{W}W_{M}^3+\hat{c}_{W} B_{M},\qquad Z_{M}= \hat{c}_{W} W_{M}^3-\hat{s}_{W} B_{M},\nonumber\\
	\hat{c}_{W}&= &\frac{g_5}{\sqrt{g_5^2+g_5^{\prime 2}}},\qquad \ \ \ \  \quad ~ ~\hat{s}_{W}= \frac{g^{\prime}_5}{\sqrt{g_5^2+g_5^{\prime 2}}},
\end{eqnarray}
with $g_5^{\prime}$ being the dimensionful 5D gauge coupling associated to the hypercharge $U(1)_Y$, given by
\begin{eqnarray}
	g_5^{\prime}= \frac{g_5 g_X}{\sqrt{g_5^2+g_X^2}}
\end{eqnarray}
in the $SU(3)$ case.~\footnote{For the holographic description of $SU(2)_1\times SU(2)_2\to SU(2)_L$ we have assumed a $P_{LR}$ symmetry $SU(2)_1\leftrightarrow SU(2)_2$ within the composite sector, leading thus to $g_{5}^{(1)}=g_5^{(2)}= \sqrt{2}g_5$ and reducing by one the number of 5D input parameters. In this case, $g_5^{\prime}$ is just the 5D gauge coupling associated to $U(1)_Y$.} Therefore, the UV boundary conditions for $Z_{\mu}$ and $W_{\mu}^{\pm}$ after EWSB read 
\begin{eqnarray}
	\left.\left(-\partial_z +\frac{v^2}{4}\left(g_5^2+g_5^{\prime 2}\right)\right)Z_{\mu}\right|_{z=R}&=&0,\\
	\left.\left(-\partial_z +\frac{v^2}{4}g_5^2\right)W_{\mu}^{\pm}\right|_{z=R}&=&0,
\end{eqnarray}
whereas all the rest remain the same. After imposing the UV boundary conditions, this leads in particular to the following KK decomposition~\footnote{Obviously, eq.~(\ref{zprime}) just holds for the $SU(3)$ case
.}
\begin{eqnarray}
	A_{\mu}(x,z)&=&\sum_n a_{\gamma}^nC(m_n^{\gamma},z)A_{\mu}^{(n)}(x),
\end{eqnarray}
\begin{eqnarray}
	Z_{\mu}(x,z)&=&\sum_n a_{Z}^n\left[C(m_n^Z,z)+\frac{1}{m_n^Z}\frac{v^2}{4}(g_5^2+g_5^{\prime 2}) S(m_n^Z,z) \right] Z_{\mu}^{(n)}(x),\label{zeta}\\
	W_{\mu}^{\pm}(x,z)&=&\sum_n a_{W}^n\left[C(m_n^W,z)+\frac{1}{m_n^W}\frac{v^2}{4}g_5^2 S(m_n^W,z) \right]W_{\mu}^{\pm(n)}(x),\label{doblew}\\
	Z_{\mu}^{\prime}(x,z)&=&\sum_n a_{Z^{\prime}}^{n}C(m_n^{Z^{\prime}},z)Z_{\mu}^{\prime(n)}(x),\label{zprime}
\end{eqnarray}
where  $C(m,z)$ and $S(m,z)$ are functions satisfying the bulk equations of motion
\begin{eqnarray}
	\left[a(z)m^2+\partial_z a(z)\partial_z\right]f(z)=0,
\end{eqnarray}
and fulfilling boundary conditions $C(m,R)=1,~ \partial_z C(m,R)=0,~S(m,R)=0,~ \partial_z S(m,R)=m$. They are given by \cite{Falkowski:2006vi, Csaki:2008zd}
\begin{eqnarray}
	C(m,z)&=&\frac{\pi}{2}mz\left[Y_0(mR)J_1(mz)-J_0(mR)Y_1(mz)\right],\\
	S(m,z)&=&\frac{\pi}{2}mz\left[Y_1(mz)J_1(mR)-J_1(mz)Y_1(mR)\right].
\end{eqnarray}

\begin{figure}[t]
\includegraphics[width=0.5\columnwidth]{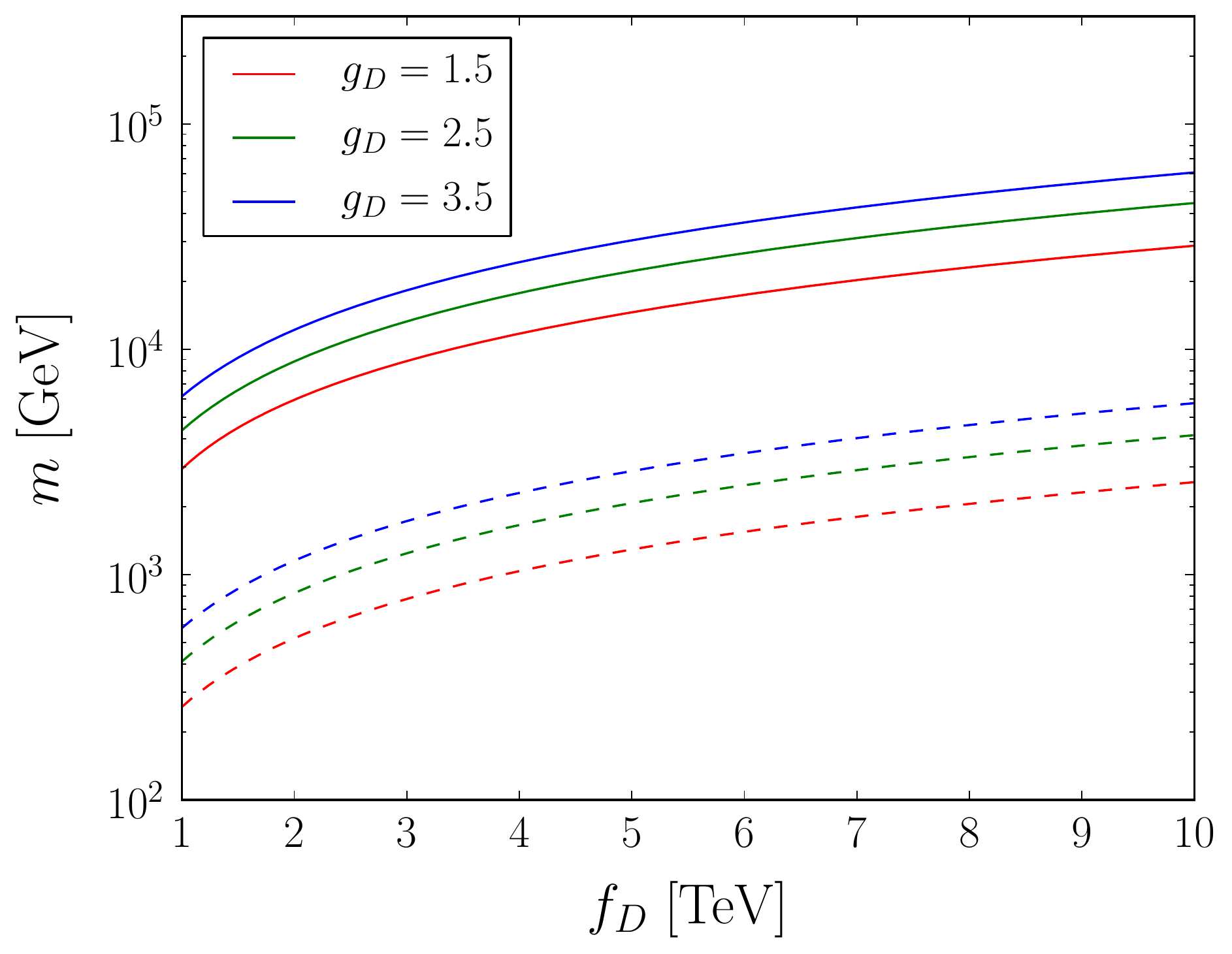}
\includegraphics[width=0.5\columnwidth]{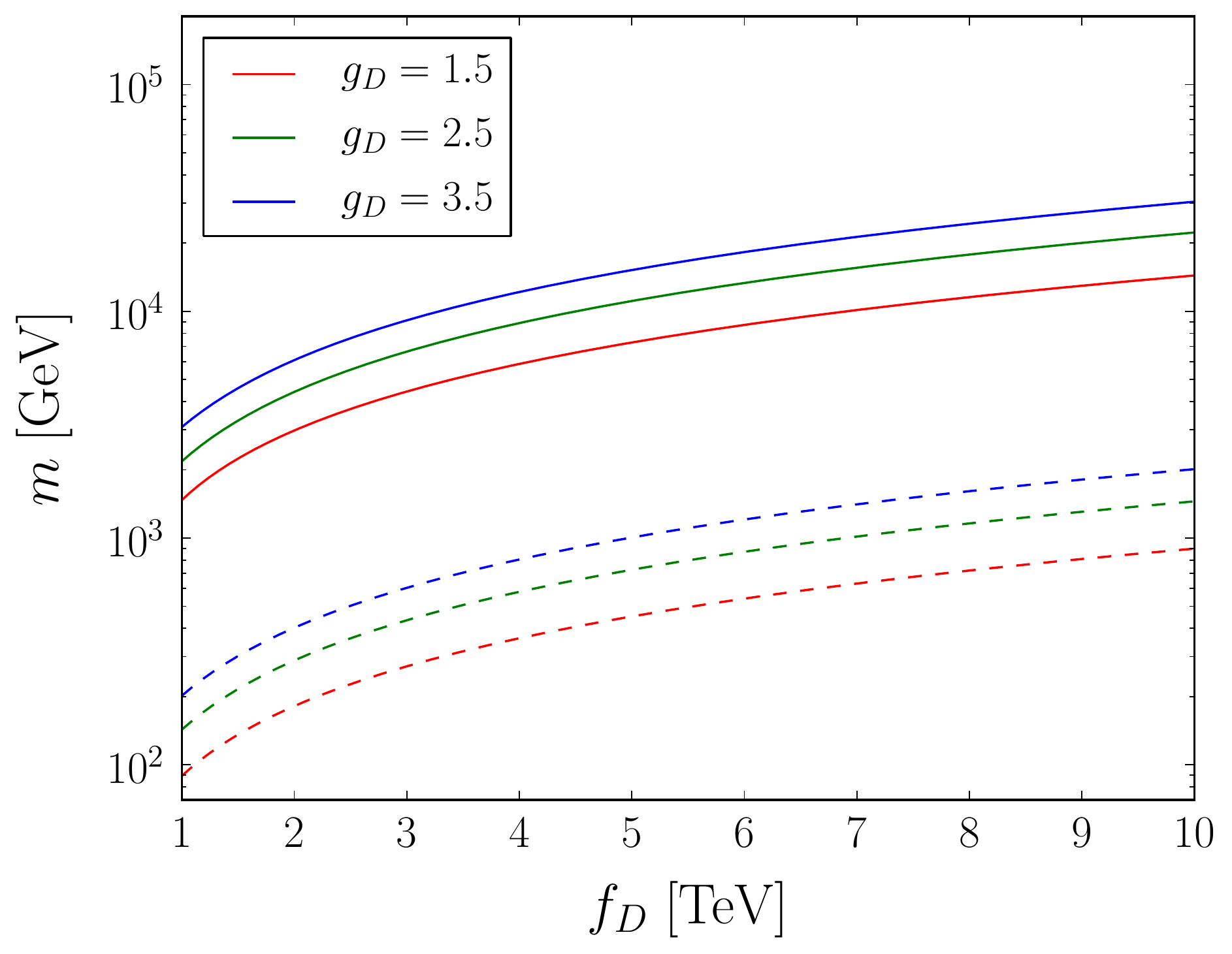}
\includegraphics[width=0.5\columnwidth]{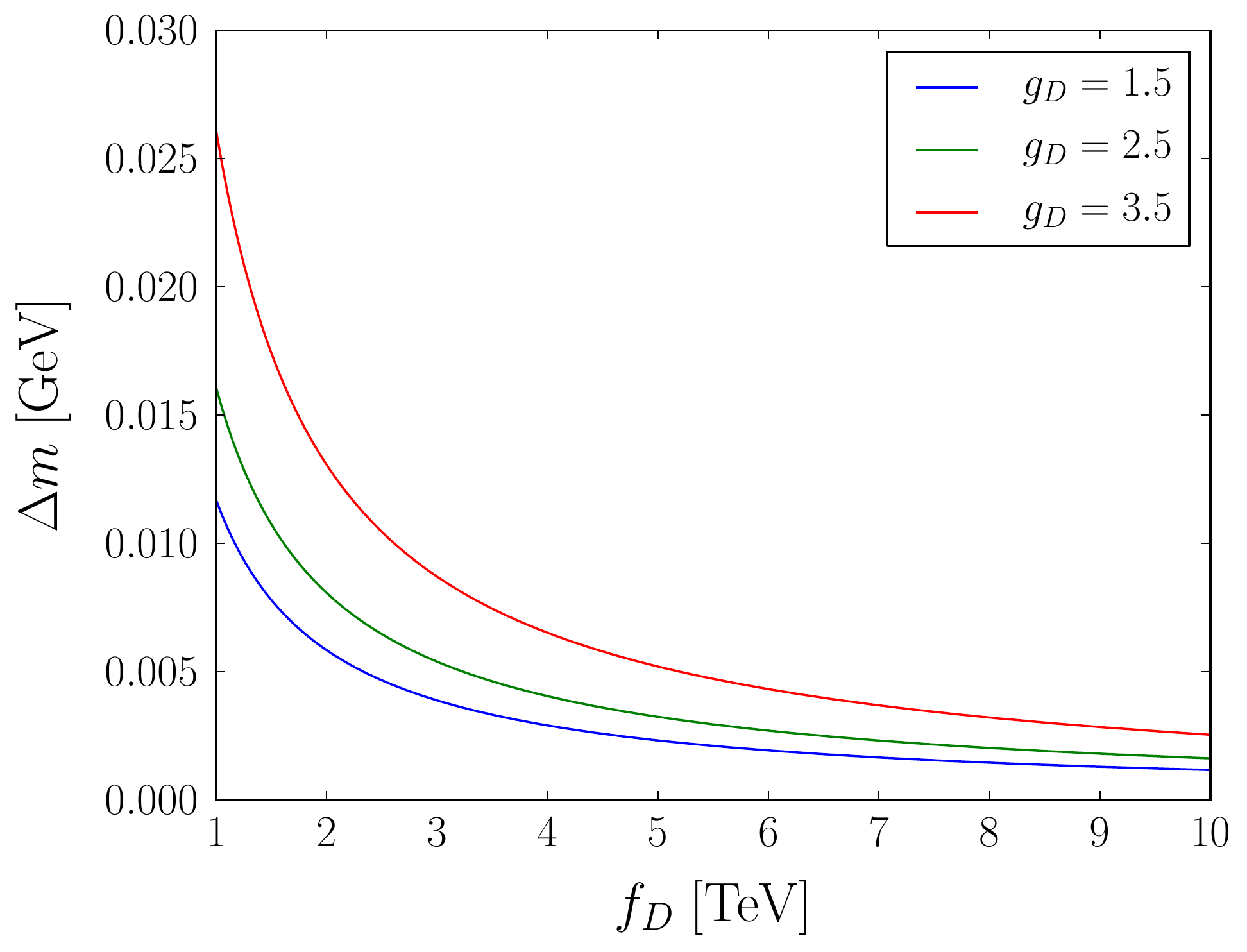}
\includegraphics[width=0.5\columnwidth]{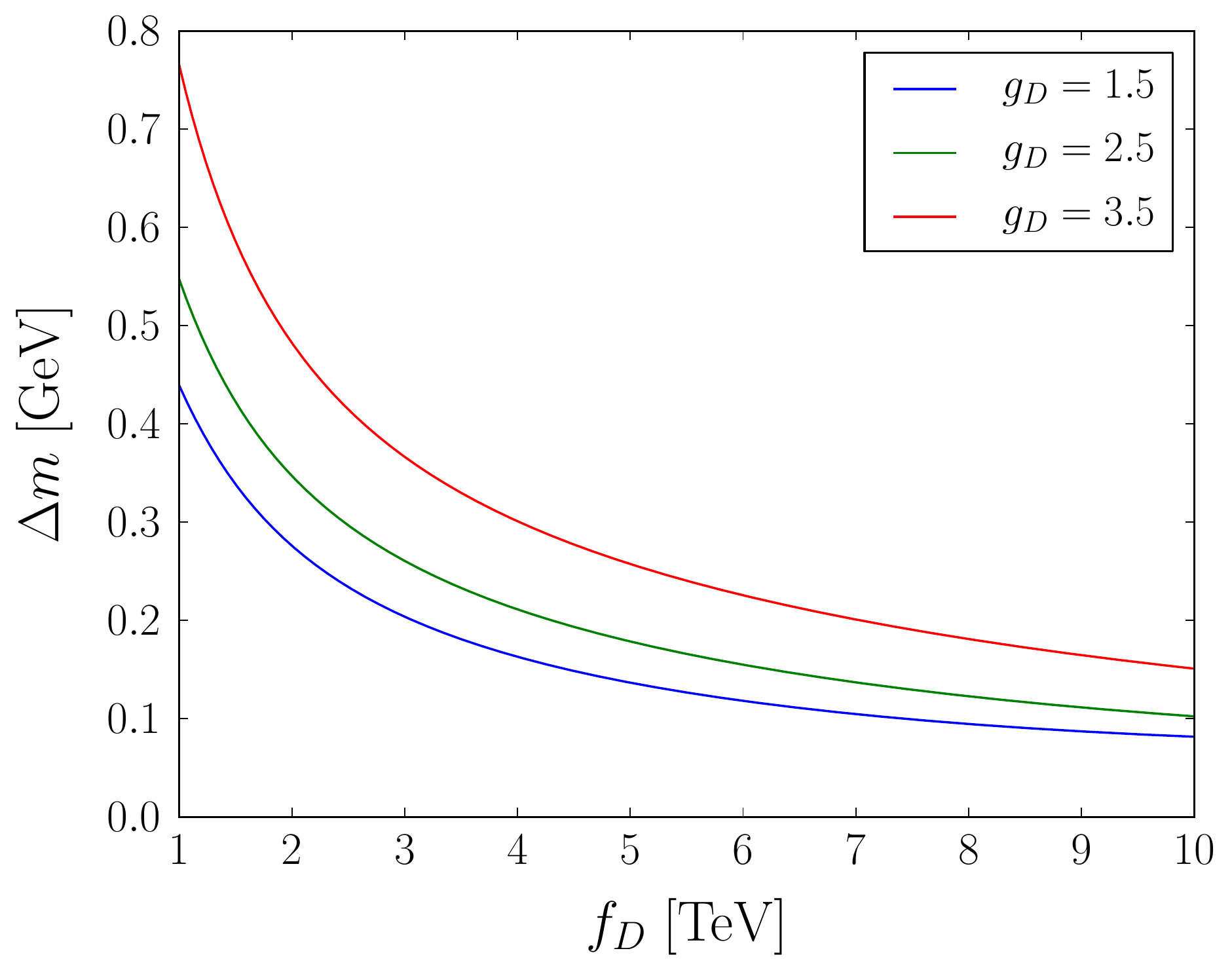}
\caption{Top left) Masses of both the charged pNGBs (dashed lines) and the heavier resonances (solid lines) as a function of $f_D$ for three different values of $g_D$ in the coset $\triplet$. Top right) Same as \textit{top left)} but for the coset $\doublet$. Bottom left) Mass difference between the neutral and the charged pNGBs as a function of $f_D$ for three diferent values of $g_D$ in the coset $\triplet$. Bottom right) Same as \textit{bottom left)} but for the coset $\doublet$.}
\label{fig:masses}
\end{figure}

It is possible to express three of the five input parameters in these 5D holographic descriptions $\{v,g_5,g_5^{\prime},R,R^{\prime}\}$ as a function of the other two by matching $G_F$, $m_Z$ and $\alpha_{\rm em}(m_Z)=e(m_Z)^2/4\pi$ to their best SM fit values. The first of these conditions just yields, 
\begin{eqnarray}
	\frac{G_F}{\sqrt{2}}=-\frac{g_5^2}{8}G_0^{(W)}(R,R), 
\end{eqnarray}
where $G_0^{(W)}(z,z^{\prime})$ is the 5D $W$ propagator evaluated at zero momentum, i.e.,
\begin{eqnarray}
	G_0^{(W)}(z,z^{\prime})=-\sum_{n=1}^{\infty}\frac{f_n^{(W)}(z)f_n^{(W)}(z^{\prime})}{m_n^{W 2}}=-\frac{4}{g_5^2v^2}-\frac{\mathrm{min}(z,z^{\prime})^2-R^2}{2R}.
\end{eqnarray}
Naively, as usual in these scenarios, one might have expected that the contribution of the KK tower would require a shift of the Higgs VEV to absorb it, i.e., $v\neq v_{\rm SM}$. However, as one can readily see using the holographic basis, all the effects arise from the UV localized operator $(D_{\mu}H)^{\dagger}D^{\mu}H$, leading to
\begin{eqnarray}
	 \frac{1}{2v_{\rm SM}^2}= \frac{G_F}{\sqrt{2}}=\frac{1}{2v^2},
\end{eqnarray}
and thus to $v\approx 246$~GeV. It is possible to get analytical expressions for the other two constraints by using the approximate expressions of $C(m,z)$ and $S(m,z)$ for $mz\le m R^{\prime}\ll 1$
\begin{eqnarray}
	C(m,z)&\approx& 1-m^2\int_R^{z}\mathrm{d}z_1a^{-1}(z_1)\int_R^{z_1}\mathrm{d}z_2~a(z_2)\nonumber\\
			 &+&m^4\int_R^{z}\mathrm{d}z_1a^{-1}(z_1)\int_R^{z_1}\mathrm{d}z_2~a(z_2)\int_R^{z_2}\mathrm{d}z_3a^{-1}(z_3)\int_R^{z_3}\mathrm{d}z_4~a(z_4)\nonumber\\
				   &=&1-\frac{1}{4}m^2R^{2}\left[1-\left(\frac{z}{R}\right)^2+2\left(\frac{z}{R}\right)^2\log\left(\frac{z}{R}\right)\right]\\
				   &+&\frac{1}{64}m^4R^4\left[1+4\left(\frac{z}{R}\right)^2-5\left(\frac{z}{R}\right)^4+4\left(\frac{z}{R}\right)^2\left(2+\left(\frac{z}{R}\right)^2\right)\log\left(\frac{z}{R}\right)\right],\nonumber
\end{eqnarray}
\begin{eqnarray}
	S(m,z)&\approx&m\int_R^z\mathrm{d}z_1 a^{-1}(z_1)-m^3\int_R^{z}\mathrm{d}z_1a^{-1}(z_1)\int_R^{z_1}\mathrm{d}z_2~a(z_2)\int_R^{z_2}\mathrm{d}z_3 a^{-1}(z_3)\nonumber\\
				   &=&\frac{mz^2}{2R}-\frac{m^3}{16R}\left(z^2-R\right)^2,
\end{eqnarray}
which leads to
\begin{eqnarray}
	m_Z\approx  \frac{1}{2}\frac{v}{\hat{c}_W} \frac{g_5}{\sqrt{R L}}\left[1-\frac{1}{32}\frac{g_5^2}{\hat{c}_W^2 R L}\frac{v^2R^{\prime 2}}{L}\right],\quad\mathrm{and}\quad  e=\frac{g_5 \hat{s}_W}{\sqrt{R L}},
\end{eqnarray}
where we have defined  for convenience the \emph{volume factor}
\begin{eqnarray}
	L= \log(R^{\prime}/R).
\end{eqnarray}
Therefore, one can write at leading order in $v^2R^{\prime 2}$
\begin{eqnarray}
	g_D= g_5 R^{-1/2}\approx g\sqrt{L}\left(1+\frac{1}{8}\frac{m_W^2 R^{\prime 2}}{c_W^2-s_W^2}\frac{1}{L}\right),\ \sin2\hat{\theta}_W\approx \sin2\theta_W\left(1-\frac{1}{8}\frac{m_Z^2 R^{\prime 2}}{L}\right),\qquad 
\end{eqnarray}
where $g_D$ is the dimensionless coupling in the strong sector. Moreover, since  we are no longer trying to address the gauge hierarchy problem, there is no need to require $1/R$ to be roughly given by the Planck scale $1/R\approx M_{\rm Pl}$. Thus, we can consider both $R$ and $R^{\prime}$ as free input parameters, or equivalently,  $g_D$ and the scale of compositeness
\begin{eqnarray}
 f_{D}= \frac{2^{0(1)}}{g_5}\left[\int_R^{R^{\prime}}\mathrm{d}z~a^{-1}(z)\right]^{-1/2}\approx \frac{2^{0(1)}\sqrt{2}}{g_D R^{\prime}}.
\end{eqnarray}
 In the above equation and hereafter $2^0$ correspond to the $SU(2)^2\times U(1)]/[SU(2)\times U(1)]$ coset whereas $2^1$ will refer to the $SU(3)/[SU(2)\times U(1)]$ one. Since $f_D$ is not a pure physical quantity and its definition is always arbitrary, we have chosen $f_D$ as the scale appearing in the argument of the $W^{\pm}$ tower contribution to the corresponding pNGB potentials (see below). This means in particular that the mass scale of the composite vector resonances $m_{\rho}$ will scale differently with $f_D g_D$ for each coset (see e.g. figure~\ref{fig:masses}).

In the UV unitary gauge, where all the NGBs of the SM are gauged away,  a scalar potential $V(h,\pi^{\Bbbk})$, with $\Bbbk=i,\hat{a}$, will be generated at the quantum level  through the interaction of the NGBs with the towers of resonances associated to the 5D gauge bosons  $\mathbb{A}_{\mu}(x,z)=A_{\mu}(x,z), Z_{\mu}(x,z), W_{\mu}^{\pm}(x,z)$ \cite{Coleman:1973jx, Antoniadis:2001cv},
\begin{eqnarray}
	V(h,\pi^{\Bbbk})=\frac{3}{2}\sum_{n=1}^{\infty}\int\frac{\mathrm{d}^4p}{(2\pi)^4}\log\left[p^2+m_n^2(h,\pi^{\Bbbk})\right],
\end{eqnarray}
where $m_n(h,\pi^{\Bbbk}),~n\in \mathbb{N},$ are the masses of all possible KK resonances $\mathbb{A}_{\mu}^{(n)}(x)$ in the presence of the background fields $\pi^{\Bbbk}$ and $h$.~\footnote{$h$ appears via UV boundary conditions and it leads to analogous expressions to (\ref{zeta}) and (\ref{doblew}) with the replacement $v\to h$, since $H=(\phi^{+},\frac{1}{\sqrt{2}}\left[h+i\phi^0\right])^T$.} The previous infinite sum can be exchanged by an integral on the Minkowski space \cite{Oda:2004rm, Falkowski:2006vi}, 
\begin{eqnarray}
	V(h,\pi^{\Bbbk})=\frac{3}{(4\pi)^2}\int_0^{\infty}\mathrm{d}k~k^3\log \rho_{h,\pi^{\Bbbk}}(-k^2),
	\label{pot}
\end{eqnarray}
where $k= \sqrt{p^2}$, and $\rho_{h,\pi^{\Bbbk}}(\omega^2),~ \omega\in\mathbb{C}$,  is some spectral function, holomorphic in the $\mathrm{Re}(\omega)>0$ part of the complex plane and with roots in the real axis encoding the spectrum of $\mathbb{A}_{\mu}(x,z)$ in the presence of the background fields $\pi^{\Bbbk}$ and $h$, i.e., 
\begin{eqnarray}
	\rho_{h,\pi^{\Bbbk}}(m_n(h,\pi^{\Bbbk}))=0,\qquad n\in \mathbb{N}.
\end{eqnarray}
Such a function will be proportional to the determinant of the linear system of equations resulting from imposing the IR boundary conditions after we remove the NGBs $\pi^{\Bbbk}$ from the bulk via the following 5D gauge transformation
\begin{eqnarray}
	\mathcal{A}_{M}(x,z)\to \Omega(z)\mathcal{A}_M(x,z)\Omega(z)^T -(i/g_5)(\partial_M\Omega(z))\Omega(z)^T,
\end{eqnarray}
where $\mathcal{A}_M\in\{L_{M}^i S_L^i, R_{M}^i S_R^i,W_{M}^a T^a\}$ (see Appendices~\ref{app:triplet} and~\ref{app:doublet} for the definition of the different generators) and
\begin{eqnarray}
	\Omega(z)= \mathrm{exp}\left(-ig_5X^{\Bbbk}\pi^{\Bbbk}\int_R^z\mathrm{d}z^{\prime} f(z^{\prime})\right).
\end{eqnarray}
The terms in the  scalar potential $V(h,\pi^{\Bbbk})$  involving $\pi^{\Bbbk}$ are expected to be finite, due to non-locality in the 5D theory \cite{Antoniadis:2001cv, vonGersdorff:2002rg}, as it is manifest by the dependence of the spectral function $\rho_{h,\pi^{\Bbbk}}$ on the Wilson line 
\begin{eqnarray}
	\mathcal{W}= \Omega(R^{\prime})=\mathrm{exp}\left(-ig_5X^{\Bbbk}\pi^{\Bbbk}\int_R^{R^{\prime}}\mathrm{d}z^{\prime} f(z^{\prime})\right)=\mathrm{exp}\left(-i2^{0(1)}f_{D}X^{\Bbbk}\pi^{\Bbbk}\right),
\end{eqnarray}
which is clearly a non-local object depending on the conformal distance between the branes. Even though UV localized terms can give infinite contributions to the $V(h)\subset V(h,\pi^{\Bbbk})$ potential, they can be renormalized in the usual way. Moreover, any $V(h)$ piece of $V(h,\pi^{\Bbbk})$ can be shifted to $V_{\rm SM}(h)$, the usual SM potential that needs to be added to (\ref{pot}).

In order to investigate the shape of the pNGB potential $V(h,\pi^{\Bbbk})$ and to be sure that the desired pattern of EWSB is not changed, we  perform an expansion of  (\ref{pot}) in powers of $h/f_D$ and $\sin(\Pi/f_D)$, where $\Pi= \sqrt{\sum_{\Bbbk}(\pi^{\Bbbk})^2}$, obtaining for the $[SU(2)^2\times U(1)]/[SU(2)\times U(1)]$ coset
\begin{eqnarray}
	V(h,\pi^{i})&\approx&\left[\lambda_{0}+\lambda_{2}\left(\frac{h}{f_D}\right)^2+\lambda_4\left\{1+\frac{1}{2}\tan^2\hat{\theta}_W \frac{\pi^{+}\pi^{-}}{\Pi^2}\right\}\left(\frac{h}{f_D}\right)^4\right]\sin^2\left(\frac{\Pi}{f_D}\right),\qquad 
	\label{su2aprox}
\end{eqnarray}
where $\pi^{\pm}=(\pi^1\mp i \pi^2)/\sqrt{2}$ and we identify  for this coset the neutral state $\pi^0=\pi^3$. We have also defined
\begin{eqnarray}
	\lambda_0&= &\frac{3}{32\pi^2}\int_{\Lambda}^{\infty}\mathrm{d}t~\frac{2 \sqrt{t} R^{\prime}}{R \bar{C}^{\prime}(\sqrt{t},R^{\prime}) \bar{S}(\sqrt{t},R^{\prime})},\label{l0}\\
	\lambda_2&= &-\frac{3}{32\pi^2}\int_{\Lambda}^{\infty}\mathrm{d}t~\frac{\bar{S}^{\prime }(\sqrt{t},R^{\prime})}{R^{\prime }\bar{C}^{\prime}(\sqrt{t},R^{\prime})^2 \bar{S}(\sqrt{t},R^{\prime})},\label{l2}
\end{eqnarray}
and
\begin{eqnarray}
	\lambda_4&=&\frac{3}{32\pi^2}\int_{\Lambda}^{\infty}\mathrm{d}t~\frac{R \bar{S}^{\prime }(\sqrt{t},R^{\prime})^2}{2 R^{\prime 3}\sqrt{t} \bar{C}^{\prime}(\sqrt{t},R^{\prime})^3 \bar{S}(\sqrt{t},R^{\prime})}.
\end{eqnarray}
\begin{figure}[t]
\includegraphics[width=0.5\textwidth]{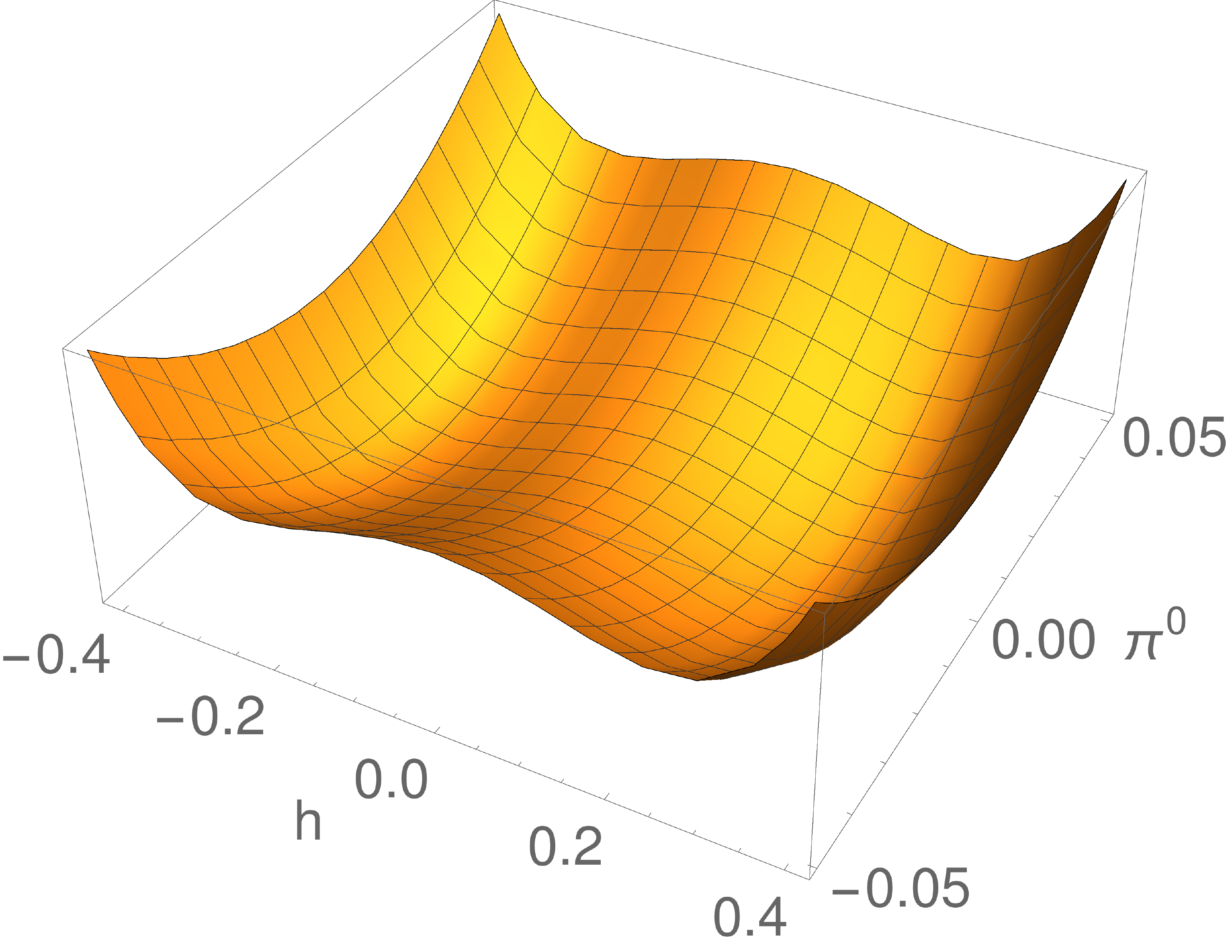}
\includegraphics[width=0.5\textwidth]{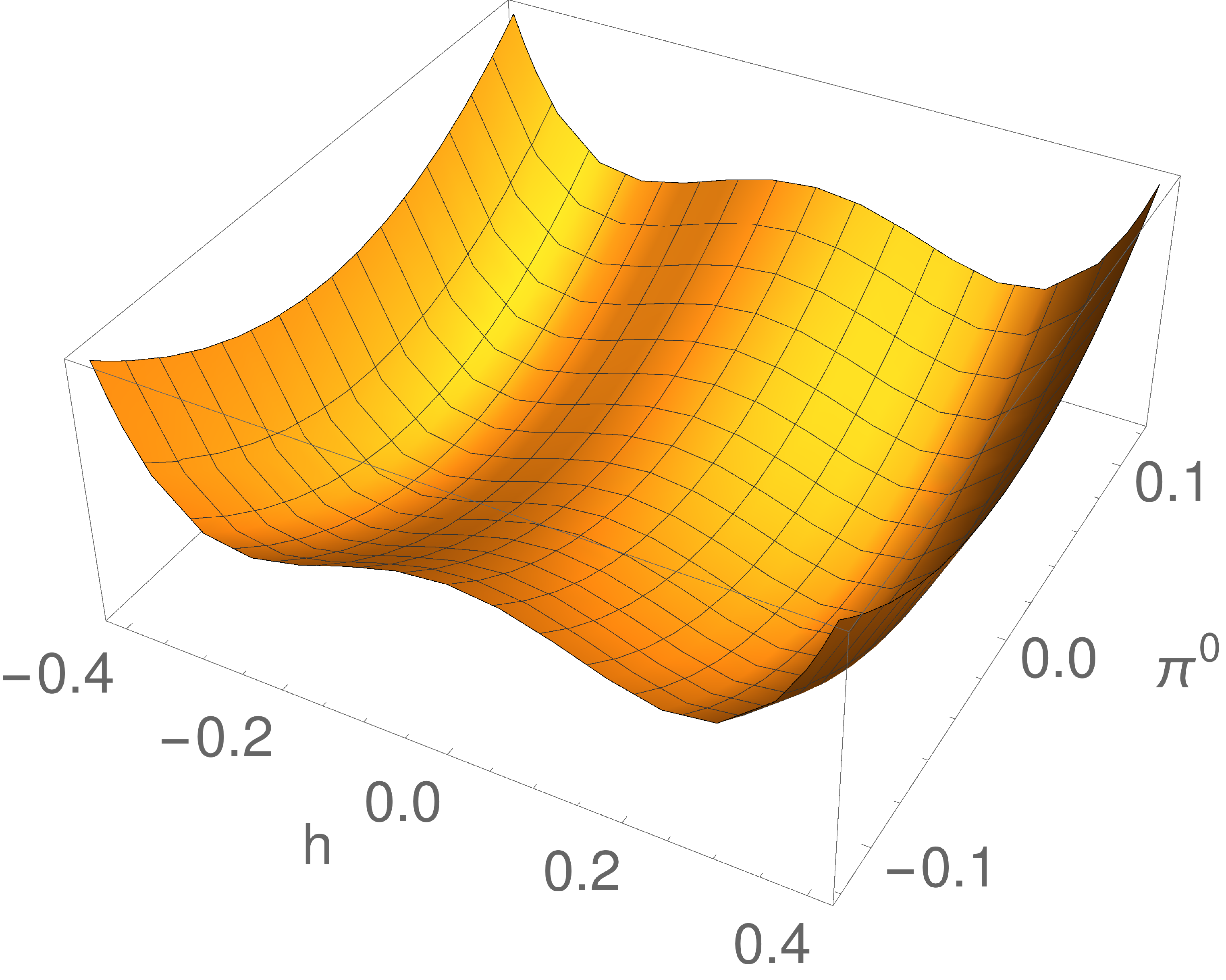}
\caption{Left) $V(h,\pi^0)+V_{\rm SM}(h)$ in the $\triplet$ coset for $f_D=1$~TeV and $g_D=3.5$, where  $\pi^{\pm}=0$.  Right) $V(h,\pi^0)+V_{\rm SM}(h)$ in the $\doublet$ coset for $f_D=1$~TeV and $g_D=3.5$, where $A^0=\pi^{\pm}=0$. }
\label{fig:potential}
\end{figure}

In the above expressions, $\bar{S}(m,z)$ and $\bar{C}(m,z)$ are the Wick-rotated versions of $S(m,z)$ and $C(m,z)$, 
\begin{eqnarray}
	\bar{C}(m,z)&= &C(i m,z)=mz\left[I_1(mz) K_0(m R)+I_0(m R) K_1(mz)\right],\\
	\bar{S}(m,z)&= &-i S(i m,z)=mz\left[I_1(mz) K_1(m R)-I_1(m R) K_1(mz)\right],
\end{eqnarray}
and $\prime$ denotes the partial derivative with respect to $z$. $\Lambda$ is an IR cut-off which has been introduced to regulate the spurious IR divergences arising from the expansion of the Coleman-Weinberg potential \cite{Csaki:2008zd, Marzocca:2012zn, Elias-Miro:2014pca}. It has been fixed by asking (\ref{su2aprox})  to reproduce the exact Goldstone mass splitting $\Delta m= m_{\pi^{\pm}}-m_{\pi^0}$. On the other hand, for the $SU(3)/[SU(2)\times U(1)]$ coset, we obtain
\begin{eqnarray}
	V(h,\pi^{\hat{a}})&\approx &\left[\lambda_{0}-\left(7+2\sec^2\hat{\theta}_W\right)\lambda_{2}\left(\frac{h}{f_D}\right)^2\right]\sin^2\left(\frac{\Pi}{f_D}\right)+\frac{1}{8}\left[\left(1+3\tan^2\hat{\theta}_W\right)\lambda_0\phantom{\frac{1}{2}}\right.\nonumber\\
																   &&+\left.\left(38-20\sec^2\hat{\theta}_W+12\sec^4\hat{\theta}_W\right)\lambda_2\left(\frac{h}{f_D}\right)^2\right]\sin^2\left(2\frac{\Pi}{f_D}\right)\nonumber\\
																   &&+~2\tan^2\hat{\theta}_W\lambda_2\left(\frac{h}{f_D}\right)^2\frac{\left((\pi^0)^2+(A^0)^2\right)^2-(\pi^{+}\pi^{-})^2}{\Pi^4}\sin^2\left(2\frac{\Pi}{f_D}\right),	   
	\label{su3aprox}
\end{eqnarray}
where now $\pi^{\pm}= (\pi^3\mp i \pi^4)/\sqrt{2}$ and we have identified for this coset $\pi^0=\pi^1$ and $A^0=\pi^2$. The couplings $\lambda_0$ and $\lambda_2$ are exactly those given by eqs.~(\ref{l0}) and (\ref{l2}) but with a different IR cut-off, reproducing the corresponding splitting $m_{\pi^{\pm}}-m_{\pi^0}= m_{\pi^{\pm}}-m_{A^0}$. 

In order to study the behavior of the scalar potential, we show in figure~\ref{fig:potential} the potential $V(h,\pi^0)+V_{\rm SM}(h)$ in both cosets, where $\pi^0$ is the lightest neutral pNGB and all other degrees of freedom have been set to zero. In both cases we have chosen benchmark values $f_D=1$~TeV and $g_D=3.5$. One can readily see from these plots that the interaction with the pNGBs does not spoil the Higgs EWSB and that they acquire no VEV as expected.

We can avoid the use of the IR cut-off introduced before by Taylor expanding (\ref{pot}) around $h=v$ and $\pi^{\Bbbk}=0$ to the renormalizable level. Even though some of the quartic self couplings arising in this expansion are still IR sensitive, all the relevant couplings for our phenomenological study are IR safe. This is indeed what we have done  to obtain all relevant couplings henceforth. 

The masses of the neutral and charged states as well as the mass difference are shown in figure~\ref{fig:masses} left and right respectively. The masses of the KK excitations are also shown for comparison. As expected, there is a large gap between the scalar and the vector resonances due to the pNGB nature of the former. In addition, we observe a rather small splitting between the neutral and the charged scalar states. This result is a consequence of the potential being loop-induced. Besides, it is worth to point out that the splitting is much smaller in $\triplet$ than in $\doublet$, because it only arises at  order  $m_W^4/f_D^4$~\footnote{There is a deep relation between this result and the fact that, in the renormalizable triplet scalar model, the corresponding quartic coupling does not renormalize under gauge interactions~\cite{Cirelli:2007xd}. } in the triplet case.

\section{Current constraints}
\label{sec:pheno}

The phenomenology of these models is completely described by only two free parameters, that we have chosen to be $g_D$ and $f_D$. We consider a region in the $f_D-g_D$ plane, parameterized by $g_D\in [1.5, 4]$ and $f_D\in [1, 10]$ TeV. This is well motivated since, as we will see, it contains a large region of parameter space for which the relic density can be exactly reproduced. In this region, restrictions coming from EW constraints on the $S, T$ and $U$ parameters are not sensitive enough to set important limits on these models, but modifications of $W$ and $Y$ do impose non-negligible bounds. These are discussed in section~\ref{sec:ewpd}. On another front, Higgs searches are not sensitive to this region of the parameter space, for both the pNGBs and the composite resonances are not light enough (see figure~\ref{fig:masses}) neither to allow the Higgs decay into invisible particles nor to modify appreciably the Higgs decay into SM fields.

Besides, direct detection experiments and monojet searches provide very weak bounds on our parameter space. Indeed, in the first case, nucleon-DM scattering processes are mediated either by loop-suppressed processes or by the t-channel exchange of a Higgs boson~\footnote{The mediation of a t-channel gauge boson results in a heavier final state, and hence suppressed by the small DM velocity.}, with a strength proportional to the coupling in the quartic $H^2 (\pi^0)^2$ term. It has been shown in previous Refs.~\cite{Frigerio:2012uc} that direct detection experiments are not sensitive to small values of this coupling, specially for large DM masses. In particular, values below $\lesssim 0.1$ are out of the reach of direct detection experiments for any DM mass. The corresponding coupling in our models turns out to be much smaller ($\approx 10^{-3}$) than this value, as can be derived from eqs. (\ref{su2aprox}) and (\ref{su3aprox}). Concerning monojet searches, the rather large pNGB masses together with the volume-suppressed couplings of the heavier vector resonances to the SM quarks put these models beyond the reach of the current LHC analyses. We have explicitly checked this observation implementing the model in \texttt{MadGraph v5}~\cite{Alwall:2014hca} by means of \texttt{FeynRules v2}~\cite{Alloul:2013bka}. We have generated Monte Carlo monojet events in the parameter space region under study. The largest cross sections correspond to the smallest values of $g_D$ and $f_D$ in the $\doublet$ model, being of order $\approx 1$ pb. We have subsequently passed these events through \texttt{Pythia v6}~\cite{Sjostrand:2006za} and implemented the CMS analysis in Ref.~\cite{Khachatryan:2014rra} (based on an integrated luminosity of 19.7 fb$^{-1}$ at $\sqrt{s} = 8$ TeV) in \texttt{MadAnalysis v5}~\cite{Conte:2014zja} for a cut on the missing energy of $\slashed{E}_T > 450$ GeV. The resulting acceptance times the production cross section after the cuts is always smaller than the upper bound on this quantity as stated by CMS, namely 7.8 fb at 95\% C.L.

Thus, beyond EWPD, the main constraints come essentially from three sides. First, from the measurement of the DM relic abundance $\Omega_\text{obs} h^2 = 0.1199\pm 0.0027$~\cite{Ade:2013zuv}, which sets an upper bound on the contribution for any DM candidate. (These bounds are discussed in section~\ref{sec:relic}.) Second, from collider searches of long-lived charged particles. Indeed, given the small splitting between the neutral and the charged scalars, the latter can be long-lived enough to scape detectors, leaving a characteristic trace because of its large mass. We consider these searches in section~\ref{sec:longlived}. (Note that these particles can be considered long-lived for collider experiment purposes, but not at cosmological scales. Thus, cosmological constraints as those coming from nucleosynthesis can be neglected~\cite{Burdin:2014xma}.) On top of that, there is a third source of constraints. These are collider bounds on the heavier resonances, the main ones coming from the LHC ATLAS and CMS Collaborations. These last constraints are detailed in section~\ref{sec:resonances}.

\begin{figure}[t]
	\hspace{-0.8cm}
\includegraphics[width=0.53\columnwidth]{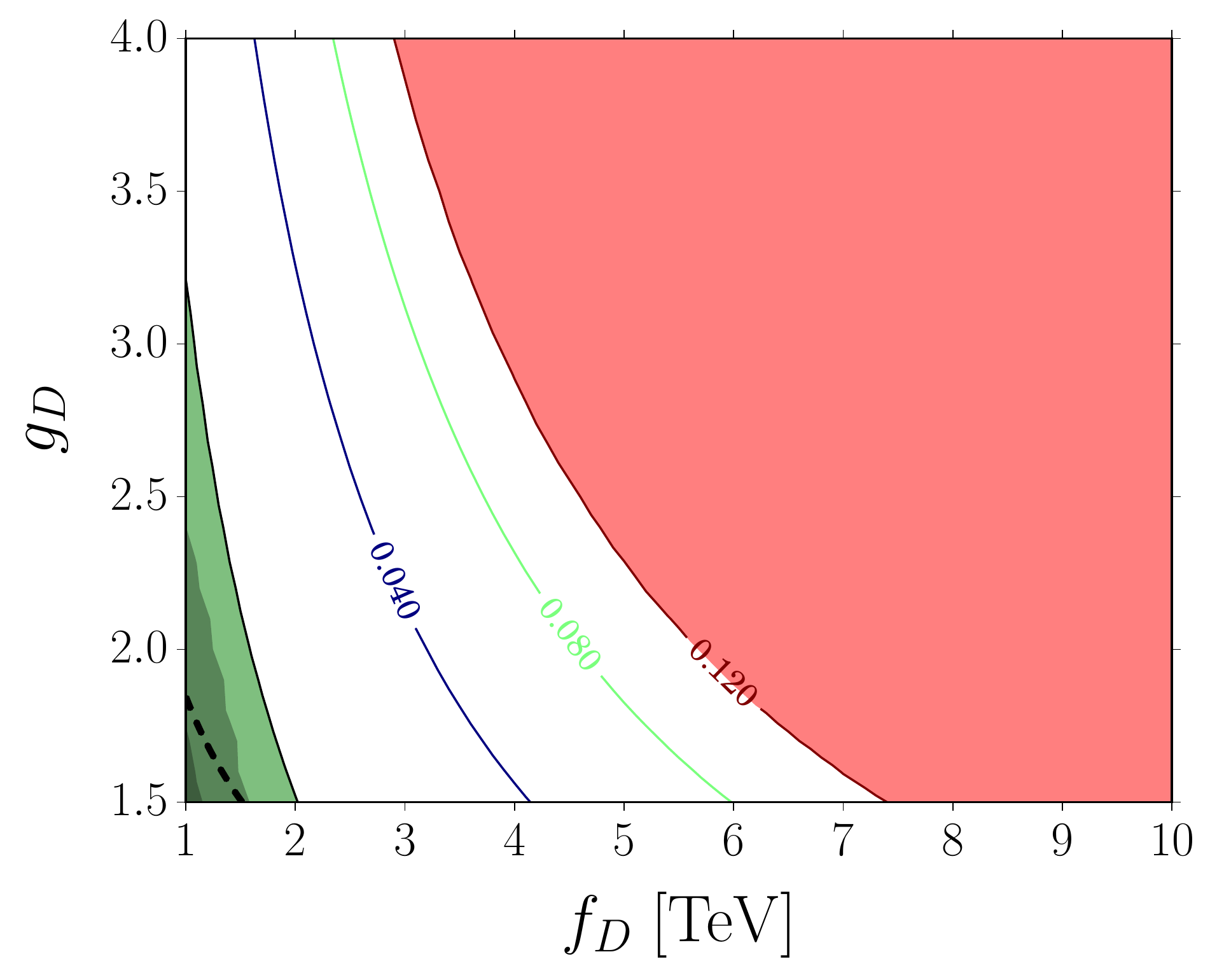}
\includegraphics[width=0.53\columnwidth]{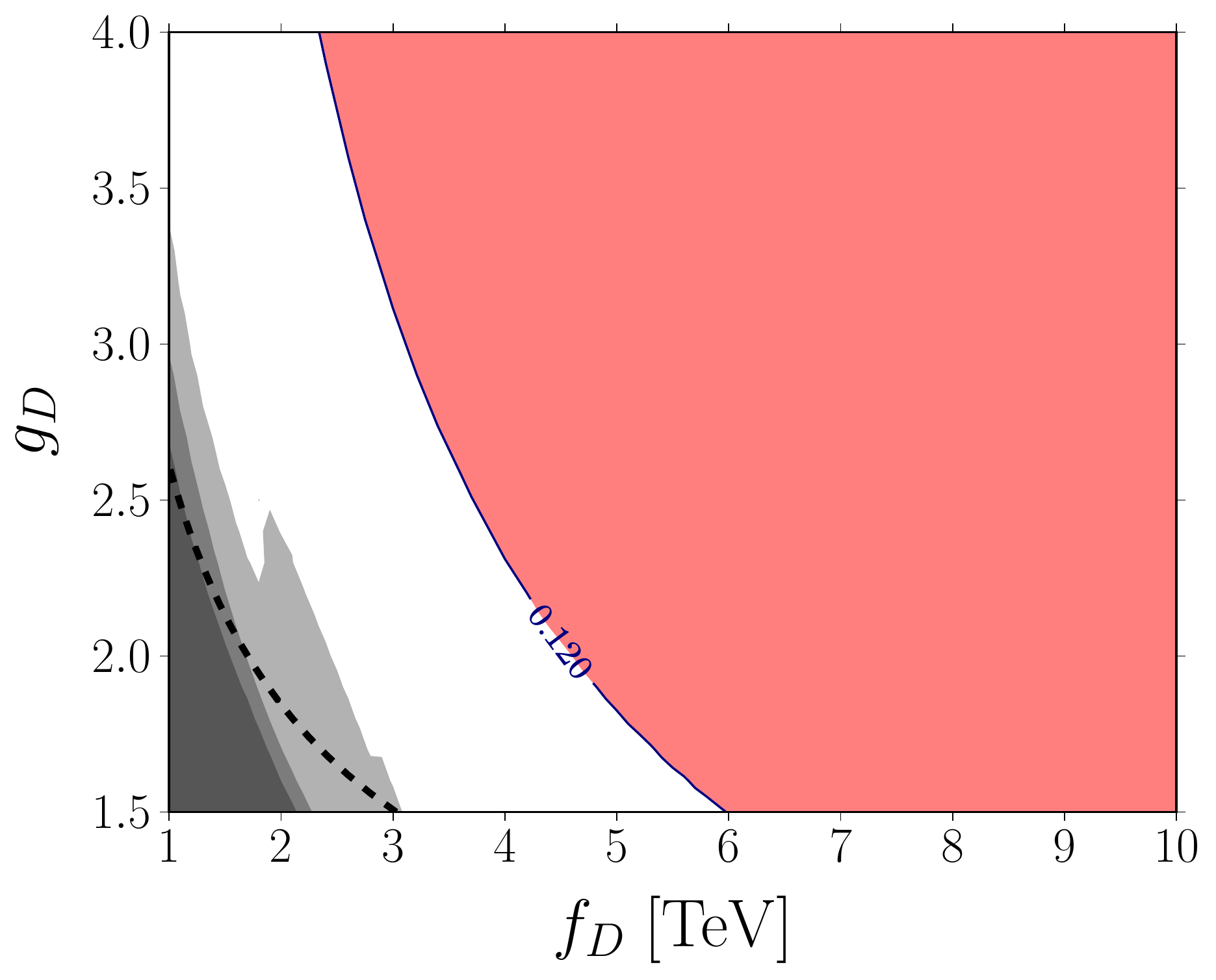}
\caption{Left) Excluded region in the coset $\triplet$ in the $f_D-g_D$ parameter space. The pink region is excluded by relic density measurements. The largest green region is excluded by searches of long-lived charged particles at the LHC. The light shaded region is excluded by dijet searches whereas the medium and dark shaded regions are excluded by $t\bar{t}$ and dilepton searches, respectively. The small dashed line encloses the region excluded by EWPD. Right) Same as \textit{Left)} but for the coset $\doublet$. }
\label{fig:excl}
\end{figure}

\subsection{Constraints from electroweak precision data}
\label{sec:ewpd}

To start with, the $T$ parameter does not receive tree-level corrections from the extra scalars, because the EW VEV is aligned in the Higgs direction, as we showed in the previous section. Besides, the loop-induced contributions from these scalars to the $S, T$ and $U$ parameters are suppressed by the splitting between the charged and the neutral components~\cite{Peskin:1991sw,Dugan:1991ck}. Hence, given that this is at least at the percent level (see figure~\ref{fig:masses}), these corrections are expected to be negligible, in light of the latest measurements: $S = -0.03\pm 0.1$, $T=0.01 \pm 0.12$ and $U =0.05 \pm 0.10$~\cite{Agashe:2014kda}. In addition, tree-level corrections to the $T$ and $S$ parameters from the heavier resonances are absent. Indeed, this can be easily understood in the dual 5D  model. If one computes the contributions to the oblique parameters arising from the integration of the KK gauge resonances, see e.g. \cite{Davoudiasl:2009cd,Carmona:2011ib}, it can be readily seen  that 
\begin{eqnarray}
	T=\frac{g^2_D\tan^2\theta_Wv^2}{4L\alpha_{\rm em}}\left[\hat{\alpha}-2\hat{\beta}+\hat{\gamma}\right]=0,\qquad S= \frac{2 g^2_D \sin^2\theta_W v^2}{L\alpha_{\rm em}}\left[-\hat{\beta}+\hat{\gamma}\right]=0,
\end{eqnarray}
since all the coefficients involved, $\hat{\alpha},\hat{\beta}$ and $\hat{\gamma}$, 
\begin{eqnarray}
	\hat{\alpha}&=&RL\int_{R}^{R^{\prime}}\mathrm{d}z~\mathrm{d}z^{\prime}a^3(z)[f_h(z)]^2\tilde{G}_0(z,z^{\prime})[f_h(z^{\prime})]^2a^3(z^{\prime}) ,\\
	\hat{\beta}&=&RL\int_{R}^{R^{\prime}}\mathrm{d}z^{\prime}~ \tilde{G}_0(R,z^{\prime})[f_h(z^{\prime})]^2a^3(z^{\prime}),\\
	\hat{\gamma}&= &RL ~\tilde{G}_{0}(R,R),\phantom{\int}
\end{eqnarray} 
become equal for a UV localized Higgs $a^3(z)[f_h(z)]^2=\delta(z-R)$, where $\tilde{G}_0(z,z^{\prime})$  is the 5D Dirichlet propagator before EWSB at zero-momentum after subtracting the corresponding zero-mode, 
\begin{eqnarray}
	\tilde{G}_0(z,z^{\prime})=\frac{z_{<}^2\left(1+2\log\left(\frac{R}{z_<}\right)\right)+z_{>}^2\left(1+2\log\left(\frac{R^{\prime}}{z_>}\right)\right)-\left[R^{\prime 2}-R^2\right]L^{-1}}{4R L},
\end{eqnarray}
and $z_{<}= \mathrm{min}(z,z^{\prime}), z_{>}= \mathrm{max}(z,z^{\prime})$.   However, we are still left with a contribution to 
\begin{eqnarray}
	W=Y=-\frac{g^2_Dv^2}{4L}\hat{\gamma}\approx \frac{g_D^2v^2}{4L}\left[\frac{1}{4}R^{\prime 2}\frac{1}{L}\right]\approx\left(\frac{g}{g_D}\right)^4\left(\frac{v}{f_D}\right)^2\frac{2^{0(1)}}{8},
\end{eqnarray}
arising from four-fermion interactions and which can be in principle non-negligible for $g_D\approx 1$ and not too large values of $f_D$. In order to study the impact of such operators on EWPD, we have performed a complete and up-to-date fit to EWPD for $S=T=0$ and $W=Y$.~\footnote{We are grateful to Jorge de Blas for providing us the $\chi^2$ for the EW fit. This fit includes all the observables considered in the analysis of~\cite{delAguila:2011zs,Blas:2013ana}, updated with the current experimental values.} The allowed values at $95\%$ C.L. are given by the region above the dashed lines in figure~\ref{fig:excl}.

\subsection{Constraints from the relic abundance observation}
\label{sec:relic}

Assuming the well-motivated standard thermal history for DM, the relic abundance can be computed using \texttt{MicrOMEGAS v4}~\cite{Belanger:2014vza}. Thus, we have also implemented the model interactions in \texttt{CalcHEP v3}~\cite{Belyaev:2012qa} by means of \texttt{FeynRules v2}. We require the computed relic density to be $\Omega h^2 \le 0.12$. The corresponding excluded region in the $f_D-g_D$ plane is shown in pink in figure~\ref{fig:excl}. In the left panel we show the results for the $\triplet$ case, while in the right panel we consider $\doublet$. Hence, the points in the frontier with the white region correspond  to the parameter space region in which the observed relic density can be explained by a single composite DM particle. Points aside this region then require extra degrees of freedom to account for the observed relic density. 

\subsection{LHC constraints on long-lived charged particles}
\label{sec:longlived}

The small splitting between the neutral and the charged states (shown in figure~\ref{fig:masses}) makes $\pi^\pm$ long-lived. Indeed, it mainly decays to the $\pi^0$~\footnote{In the doublet case,  $\pi^{\pm}$ could decay also into $A^0$  provided that $m_{A^{0}}\le m_{\pi^{\pm}}$.} through the emission of an off-shell $W^\pm$ gauge boson, being the total width given by the approximate expression~\cite{delAguila:2013aga}
\begin{equation}\label{eq:lifetime}
\Gamma \approx \frac{g^4\alpha}{48 \pi^3}\frac{\Delta m^5}{m_W^4},
\end{equation}
where $\alpha = 1$ $(1/2)$ in the triplet (doublet) case. In $\triplet$, the lifetime $\tau = 1/\Gamma$ is large enough to allow $\pi^\pm$ to scape the detectors, if it is produced in high-energy collisions. In $\doublet$, on the contrary, the decay always takes place in the detectors. The decay products are however too soft to trigger the corresponding final state in order to make it emerge from the huge $W$+ jets background. However, in the triplet case, the trace of these particles can be still observed for they give rise to anomalous energy loss. In fact, there are dedicated analyses to search for this kind of signatures. In particular, the CMS Collaboration has reported the latest constraints in Ref.~\cite{Chatrchyan:2013oca}. To our knowledge, these are the strongest limits from collider experiments. In that article, bounds are provided for different type of charged states. Among them, we can find limits for pair-produced staus, which can be directly translated to our model if the theoretical cross section for the pair-produced charged scalars is computed. In order this cross section to be calculated, we again use \texttt{MadGraph v5}. In figure~\ref{fig:excl}, the green region below the solid black line represents the parameter space points for which the cross section exceeds the values provided by CMS.

\subsection{LHC constraints on new heavy resonances}
\label{sec:resonances}

Heavy gauge boson partners are common predictions in models with composite sectors (or in their extra-dimensional dual models). In our case, as a consequence of having elementary SM fermions, these resonances will couple to them with universal and volume-suppressed couplings, i.e., $\approx g/g_D$. Moreover, for the same reason, the interactions with the SM Higgs will be also volume suppressed whereas those to the pNBGs will be on the contrary volume enhanced.  In particular, this means  that unless $g/g_D\approx 1$ they will have an important fraction of invisible decays making their collider observation extremely challenging even for small values of $f_D$. Thus, LHC constraints on these resonances will just be relevant for small values of both $g_D$ and $f_D$. Indeed, these vector resonances are expected to be heavy and narrow, what allows us to directly translate the limits on new resonances provided in the LHC analyses to our model once the corresponding cross sections are computed. Among these analyses, we consider searches of: \textit{(i)} heavy resonances decaying into pair of jets. We take the limits from the CMS analysis in Ref.~\cite{Khachatryan:2015sja}, in which the data set corresponds to a collected luminosity of 19.7 fb$^{-1}$ at $\sqrt{s} = 8$ TeV. The points excluded by this analysis are given by the light shaded region in figure~\ref{fig:excl}; \textit{(ii)} searches of $t\overline{t}$ resonances. We focus on the analyses provided by Ref.~\cite{Chatrchyan:2013lca}. The center of mass energy and integrated luminosity are the same as before. Given the small branching ratio into $t\overline{t}$, the bounds are much weaker. These are shown in figure~\ref{fig:excl} in the medium dark shaded region; \textit{(iii)} searches of dilepton resonances. We use the latest limits provided in Ref.~\cite{Aad:2014cka}. The corresponding excluded region in figure~\ref{fig:excl} is given by the dark shaded region. There is a last analysis that could be potentially useful, namely the search of pair-produced heavy resonances, as those provided in~\cite{ATLAS:2012hpa} and~\cite{Khachatryan:2014lpa}. However, their sensitivity to large resonance masses is still very limited, and so the bounds turn out to be completely negligible in the region of parameter space under consideration. 

\section{Conclusions}
\label{sec:conclusions}
\begin{figure}[t]
\begin{center}
\includegraphics[width=\textwidth]{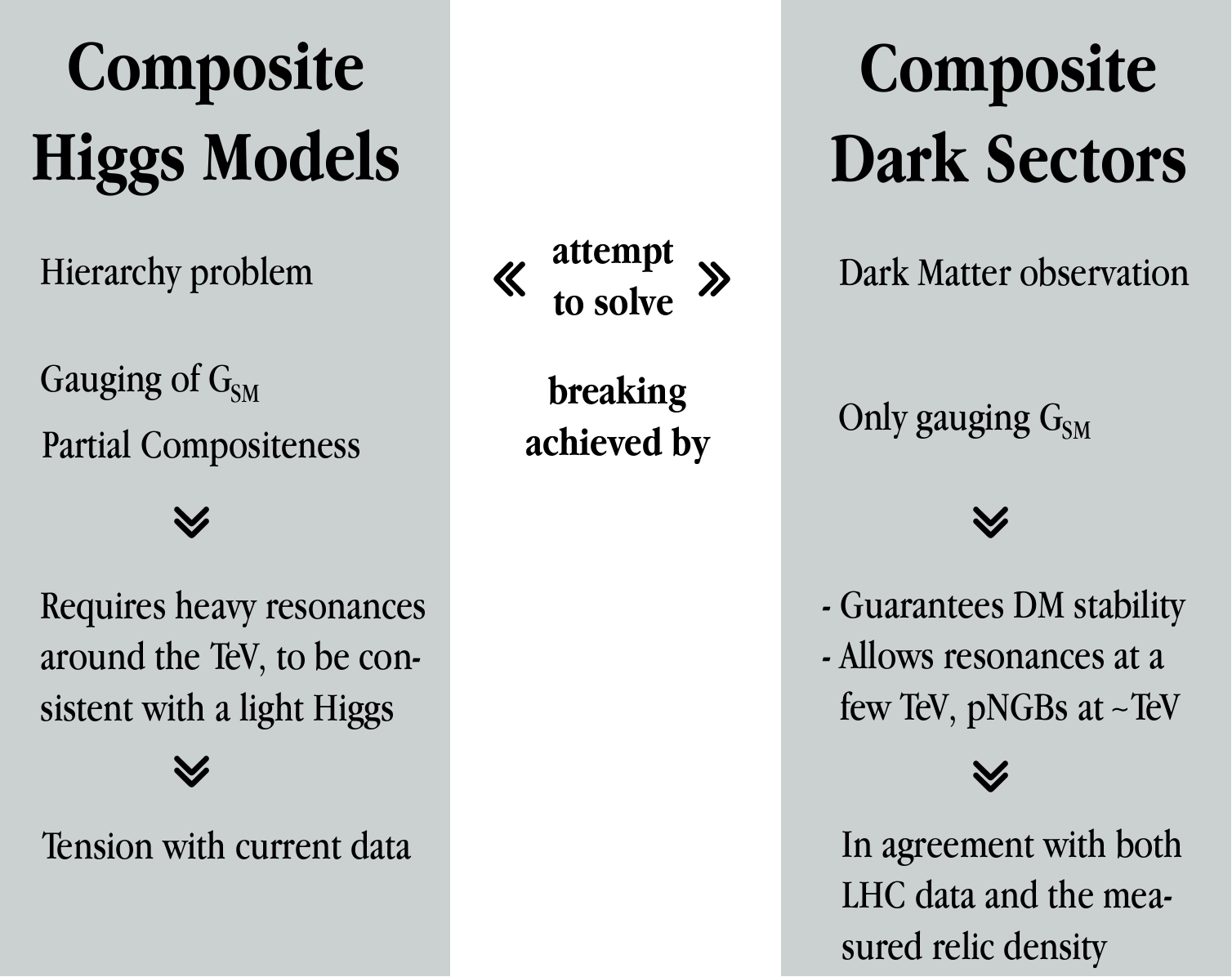}
\caption{Schematic representation of the Composite Higgs model versus the Composite Dark Sector formalism.}
\label{fig:scheme}
\end{center}
\end{figure}
We have presented a novel solution to the dark matter (DM) problem, provided by a weakly interacting composite massive particle. This solution is based on using composite pseudo Nambu-Goldstone bosons not to solve the hierarchy problem but to provide a good DM candidate.~\footnote{Similar conceptual approaches have been previously considered in~\cite{Kilic:2009mi,Kilic:2010et,Antipin:2014qva, Antipin:2015xia}.} Indeed, as depicted in figure~\ref{fig:scheme}, once we shift the focus from the hierarchy problem to the DM explanation (allowing the Higgs boson to be fully elementary too), then the minuses in Composite Higgs Models (CHMs) become pros in our new setup. In particular, CHMs require the Standard Model (SM) fermions to mix linearly with fermionic resonances in order to achieve the electroweak symmetry breaking (EWSB). It is well known that this mechanism can only provide a light Higgs boson in minimal models if the composite resonances are $\lesssim$ TeV, which is more and more in tension with the current LHC data. As a matter of fact, the latest searches of vector-like quarks~\cite{ATLAS:2015fka} exclude top partners with masses below $750$ GeV in a complete model-independent way, and even masses around $\approx 950$ GeV can be reached depending of the decay mode. This is not longer true in Composite Dark Sectors, where the DM particles are allowed to be heavy enough to account for the relic density measurement, thus  setting the composite resonances $\gtrsim$ few TeV and hence not in conflict with the current LHC data. Besides, the masses (and in fact the whole potential) of the scalar DM particles are only induced by loops of SM gauge bosons, what guarantees the DM stability even after EWSB.

In order to quantitatively discuss all these features, we have considered in detail two minimal realizations of Composite Dark Sectors, corresponding to the cosets $\triplet$ and $\doublet$, which give rise to a real scalar triplet with $Y=0$ and a complex scalar doublet with $Y=1/2$, respectively. In section~\ref{sec:potential} we have worked out the gauge interactions as well as the induced Coleman-Weinberg potential. For such a purpose, we consider a dual description of the composite sector with modified boundary conditions with respect to CHMs. We have explicitly shown that the minimum of the DM potential is always aligned in the direction that preserves the EW symmetry. We have also computed the masses of the neutral and charged scalars and the composite vector resonances as a function of the only two free parameters in the model, namely $f_D$ and $g_D$, corresponding to the typical scale and coupling of the composite sector. These two simple realizations turn out to correctly describe the DM phenomenology. We have considered experimental constraints from relic density measurements, direct detection experiments, EW precision data ---including the effects of both the scalars and the heavier resonances on the oblique parameters---, LHC searches of long-lived charged particles, collider searches of heavy narrow resonances ---dijet, $t\bar{t}$ and dilepton final states--- and LHC searches of monojet events. All together, they bound a large region of the $f_D-g_D$ plane, while allowing the dark sector to be the single component of the DM content in the universe (see figure~\ref{fig:excl}). As it has been previously pointed out (see e.g.~\cite{Chala:2015ama}), there is a significant complementarity between collider and non-collider searches, that will certainly allow us to probe the complete parameter space region in near future experiments.  In addition, these models can be disentangled from other DM solutions. In particular, two main predictions of our scenario are the presence of long-lived charged
scalars in the mass region $300\,\text{GeV} \lesssim m_{\pi^{\pm}} \lesssim 2000$ GeV and heavy narrow vector resonances decaying equally into all the SM fermions. These signatures contrast with those of other related models of DM previously considered in the literatute~\cite{Cirelli:2005uq, Khlopov:2008ty,  Hamaguchi:2009db, Frigerio:2012uc, Chala:2012af, Heikinheimo:2013fta, Cline:2013zca, Antipin:2014qva, Yamanaka:2014pva, Antipin:2015xia}. A dedicated study of the reach of future experiments to distinguish among these different alternatives is
however beyond of the scope of this paper, and will be considered elsewhere.

\acknowledgments It is a pleasure to thank Alejandro Ibarra, Felix Kahlhoefer, Jose Santiago and Pedro Schwaller for useful discussions. Special thanks to Jorge de Blas for his help with the fit to  electroweak precision data. M.C. thanks ITP at ETH Z\"urich for hospitality during the completion of this work. A.C. is supported by the Swiss National Science Foundation under contract SNSF 200021-143781.

\appendix

\section{Group theory of $\triplet$}
\label{app:triplet}

Since the pNGBs are not charged under the unbroken $U(1)$ (identified with the hypercharge), the relevant group theory is actually encoded in $SU(2)\times SU(2)$. The three generators of the left $SU(2)$ can be chosen to be
\begin{equation*}
S_L^1 = \begin{pmatrix} 0 & 0 & 0 & -\frac{i}{2}\\ 0 & 0 & -\frac{i}{2} & 0\\ 0 & \frac{i}{2} & 0 & 0\\ \frac{i}{2} & 0 & 0 & 0\end{pmatrix}, \qquad S_L^2 = \begin{pmatrix}0 & 0 & \frac{i}{2} & 0\\ 0 & 0 & 0 & -\frac{i}{2}\\ -\frac{i}{2} & 0 & 0 & 0\\ 0 & \frac{i}{2} & 0 & 0\end{pmatrix}, \qquad S_L^3 = \begin{pmatrix}0 & -\frac{i}{2} & 0 & 0\\ \frac{i}{2} & 0 & 0 & 0\\ 0 & 0 & 0 & -\frac{i}{2}\\ 0 & 0 & \frac{i}{2} & 0\end{pmatrix};
\end{equation*}
while the right $SU(2)$ is generated by
\begin{equation*}\label{eq:tripletgenerators}
S_R^1 = \begin{pmatrix} 0 & 0 & 0 & \frac{i}{2}\\ 0 & 0 & -\frac{i}{2} & 0\\ 0 & \frac{i}{2} & 0 & 0\\ -\frac{i}{2} & 0 & 0 & 0\end{pmatrix}, \qquad S_R^2 = \begin{pmatrix}0 & 0 & \frac{i}{2} & 0\\ 0 & 0 & 0 & \frac{i}{2}\\ -\frac{i}{2} & 0 & 0 & 0\\ 0 & -\frac{i}{2} & 0 & 0\end{pmatrix}, \qquad S_R^3 = \begin{pmatrix}0 & -\frac{i}{2} & 0 & 0\\ \frac{i}{2} & 0 & 0 & 0\\ 0 & 0 & 0 & \frac{i}{2}\\ 0 & 0 & -\frac{i}{2} & 0 \end{pmatrix}.
\end{equation*}
Indeed, for any $i,j$, $[S_L^i, S_R^j] = 0$ and $[S_{L(R)}^i,S_{L(R)}^j] = i \epsilon_{ijk} S^k_{L(R)}$, where $\epsilon_{ijk}$ stands for the totally antisymmetric tensor. The unbroken subgroup $SU(2)_V$ is generated by the linear combinations $T^i = S_L^i + S_R^i$. The electromagnetic operator is then given by $Q = S_3$. The coset generators are written as $X^i = S_L^i - S_R^i$.

\section{Group theory of $\doublet$}
\label{app:doublet}

In standard notation, the $SU(3)$ generators are written as
\begin{equation*}\label{eq:doubletgenerators}
S^1 = \begin{pmatrix}0 & \frac{1}{2} & 0\\ \frac{1}{2} & 0 & 0\\ \w 0 &\w 0 &\w 0\end{pmatrix}, \quad S^2 = \begin{pmatrix} 0 & \frac{i}{2} & 0\\ \frac{i}{2} & 0 & 0\\ \w 0 &\w 0 &\w 0\end{pmatrix}, \quad S^3 = \begin{pmatrix} \frac{1}{2} & 0 & 0\\ 0 & -\frac{1}{2} & 0\\ \w 0 &\w 0 &\w 0\end{pmatrix}, \quad S^4 = \begin{pmatrix} 0 & 0 & \frac{1}{2}\\ \w 0 &\w 0 &\w 0\\ \frac{1}{2} & 0 & 0\end{pmatrix},
\end{equation*}
\begin{equation*}
S^5 = \begin{pmatrix} 0 & 0 & -\frac{i}{2}\\\w 0 &\w 0 &\w 0\\ \frac{i}{2} & 0 & 0\end{pmatrix}, \quad S^6 = \begin{pmatrix}\w0 &\w 0 &\w 0\\ 0 & 0 & \frac{1}{2}\\ 0 & \frac{1}{2} & 0 \end{pmatrix}, \quad S^7 = \begin{pmatrix} \w0 &\w 0 &\w 0\\ 0 & 0 & -\frac{i}{2}\\ 0 & \frac{i}{2} & 0\end{pmatrix}, \quad S^8 = \frac{1}{\sqrt{3}}\begin{pmatrix} \frac{1}{2} & 0 & 0\\ 0 & \frac{1}{2} & 0\\ \w 0 &\w 0 & -1\end{pmatrix}.
\end{equation*}
The matrices $T^1 = S^1, T^2 = S^2$ and $T^3 = S^3$ generate the unbroken $SU(2)$ subgroup, while $U(1)$ is generated by $S^8/\sqrt{3}$. The electromagnetic charge is thus written as $Q = S^3 + S^8/\sqrt{3}$. Finally, the coset generators are given by $X^1 = S^4, X^2 = S^5, X^3 = S^6$ and $X^4 = S^7$.

\bibliographystyle{JHEP_improved}
\bibliography{biblio}{}

\end{document}